\documentclass[preprint,preprintnumbers,jcp,showpacs,superscriptaddress]{revtex4-1}

\usepackage{palatino}
\usepackage[latin1]{inputenc}
\usepackage{epsf}
\usepackage{amsmath}
\usepackage{amssymb}
\usepackage{latexsym}
\usepackage{calc}
\usepackage{color}
\usepackage{shadow}
\usepackage{epsfig}
\usepackage[pdftex, pdfstartview = {FitH}]{hyperref}
\usepackage{graphicx}
\usepackage{multirow} %added
\usepackage{ulem}

\def\br{{\bf r}}
\def\bR{{\bf R}}

\def\bA{{\bf A}}

\newcommand{\ul}[1]{\underline{#1}}
\newcommand{\dul}[1]{\underline{\underline{#1}}}
\newcommand{\Am}{$\mathcal A$-matrix}

\begin{document}
\title{On the mass of atoms in molecules: Beyond the Born-Oppenheimer approximation}

\author{Arne Scherrer}
\affiliation{Martin-Luther-Universit\"at Halle-Wittenberg, von-Danckelmann-Platz 4, D-06120 Halle, Germany}
\affiliation{UMR 8640 ENS-CNRS-UPMC, D\'epartement de Chimie, 24 rue Lhomond, \'Ecole Normale Sup\'erieure, 75005 Paris, France}
\affiliation{UPMC Universit\'e Paris 06, 4, Place Jussieu, 75005 Paris, France}
\author{Federica Agostini}
\email{agostini@mpi-halle.mpg.de}
\affiliation{Max-Planck-Institut f\"ur Mikrostrukturphysik, Weinberg 2, D-06120 Halle, Germany}
\author{Daniel Sebastiani}
\affiliation{Martin-Luther-Universit\"at Halle-Wittenberg, von-Danckelmann-Platz 4, D-06120 Halle, Germany}
\author{E. K. U. Gross}
\affiliation{Max-Planck-Institut f\"ur Mikrostrukturphysik, Weinberg 2, D-06120 Halle, Germany}
\author{Rodolphe Vuilleumier}
\email{rodolphe.vuilleumier@ens.fr}
\affiliation{UMR 8640 ENS-CNRS-UPMC, D\'epartement de Chimie, 24 rue Lhomond, \'Ecole Normale Sup\'erieure, 75005 Paris, France}
\affiliation{UPMC Universit\'e Paris 06, 4, Place Jussieu, 75005 Paris, France}

\begin{abstract}
Describing the dynamics of nuclei in molecules requires a potential energy surface, which is traditionally provided by the Born-Oppenheimer or adiabatic approximation. However, we also need to assign masses to the nuclei. There, the Born-Oppenheimer picture does not account for the inertia of the electrons and only bare nuclear masses are considered. Nowadays, experimental accuracy challenges the theoretical predictions of rotational and vibrational spectra and requires to include the participation of electrons in the internal motion of the molecule. More than 80 years after the original work of Born and Oppenheimer, this issue still is not solved in general. Here, we present a theoretical and numerical framework to address this problem in a general and rigorous way. Starting from the exact factorization of the electron-nuclear wave function, we include  electronic effects beyond the Born-Oppenheimer regime in a perturbative way via position-dependent corrections to the bare nuclear masses. This maintains an adiabatic-like point of view: the nuclear degrees of freedom feel the presence of the electrons via a single potential energy surface, whereas the inertia of electrons is accounted for and the total mass of the system is recovered. This constitutes a general framework for describing the mass acquired by slow degrees of freedom due to the inertia of light, bounded particles. We illustrate it with a model of proton transfer, where the light particle is the proton, and with corrections to the vibrational spectra of molecules. Inclusion of the light particle inertia allows to gain orders of magnitude in accuracy.
\end{abstract}

\maketitle
%\tableofcontents
\clearpage

\section{Introduction}
The Born-Oppenheimer (BO)~\cite{BO}, or adiabatic, treatment of the coupled motion of electrons and nuclei in molecular systems is among the most fundamental approximations in condensed matter physics and chemical physics. Based on the hypothesis that part of the system, usually electrons or protons, evolves on a much shorter time-scale than the rest, i.e. (heavy) nuclei or ions, the BO approximation allows one to visualize molecules as a set of nuclei moving on a single potential energy surface that represents the effect of the electrons in a given eigenstate. Yet, it is an approximation, yielding the correct dynamics only in the limit of infinite nuclear masses. Consequently when compared to highly accurate molecular spectroscopy measurements, theoretical predictions might deviate from experimentally observed behavior. 

In those situations, the question of which \textsl{masses}~\cite{Lee_CPL1996, Kutzelnigg_MP1997, Kutzelnigg_MP2007, Goldhaber_PRA2005} are to be considered when calculating rotational and vibrational spectra of light molecules, for instance hydrogen-based~\cite{Moss_MP1977, Moss_MP1977_2, Schwenke_JPCA2001, Tennyson_JCP1999, Tennyson_JCP2015, Wolf_PRL2012, Jow-Tsong_PRA2013}, often appears in the literature to rationalize this problem. In the BO approximation, the electrons appear only implicitly in the dynamics, as a potential energy contribution to the Hamiltonian driving the motion of the nuclei. The kinetic energy arising from the molecular motion then involves only the bare nuclear masses. However, electrons are carried along with the nuclei, thus how is their inertia accounted for? It has been proposed that more accurate results are obtained when employing \textsl{atomic masses} rather than bare nuclear masses~\cite{Csaszar_JCP2012}. 

The measured ro-vibrational spectrum of hot water in sunspots, for example, is very dense, with about 50 lines per wavenumber~\cite{Wallace1995}. However, their assignment can not be performed at the BO level, either using nuclear or atomic masses, because of the lack of accuracy. Adding half an electron mass to the proton to effectively include non-adiabatic effects has been shown to lead to better results~\cite{Polyansky1997}. Such \textsl{fractional masses} account for the bond ionicity but there is no systematic way to include such corrections.

One solution to the problem is to perform a full non-adiabatic treatment of the coupled electron-nuclear problem, but the numerical cost is much larger than a BO calculation. Also, from a fundamental point of view, this does not answer the question of what is the mechanism by which the inertia of the electrons affects the mass of the heavy degrees of freedom. An alternative approach, pioneered by Bunker and Moss~\cite{Moss_MP1977, Moss_MP1977_2, Moss_JMS1980}, is to treat perturbatively non-adiabatic effects, but applications are still limited to di- and tri-atomic molecules. In connection to the perturbation idea of Bunker and Moss, accurate numerical calculations have been performed on small molecules, like H$_2$, D$_2$, HD, H$_3^+$~\cite{Wolniewicz_JCP1995, Komasa_JCP2009, Csaszar_JCTC2013, Csaszar_JCP2014, Tennyson_PRA2013}. However, despite the effort to push forth the applications, it seems that the basic formalism still represents a major obstacle for the treatment of molecular systems comprising more than three atoms. The main reason is to be found in the use of internal coordinates, obtained after separation of the rotational and vibrational degrees of freedom of the center of mass of the molecule, as starting point for the application of the perturbation approach. In internal coordinates, the Hamiltonian of the molecular system is usually only handled numerically already for the tri-atomic case and difficulties are encountered when trying to rationalize the outcome of the computation.

In the present paper we examine this problem in the framework of the exact factorization of the electron-nuclear wave function~\cite{Gross_PRL2010}. This (non-adiabatic) reformulation of the quantum-mechanical problem is used as a starting point to develop a procedure that settles the issue described above in a rigorous way. The key point in the exact factorization is that the electronic effect on the nuclear system is taken into account by time-dependent vector and scalar potentials. These concepts are the generalization of similar, but static, quantities appearing also within the BO approximation. We show that non-adiabatic effects can be accounted for, by formulating a theory that treats these effects as a perturbation to the BO problem. Such a framework has been discussed in previous work~\cite{Scherrer_JCP2015} to derive the nuclear velocity perturbation theory~\cite{Scherrer_JCTC2013} for vibrational circular dichroism~\cite{Nafie_JCP1983}. As we will show below, here we propose a new perspective on the nuclear velocity perturbation theory, which will allow us to access a broader class of both static, e.g. energetics, and dynamical, e.g. vibrational spectra, problems in quantum mechanics. Within the nuclear velocity perturbation theory, non-adiabatic effects can be included by taking into account corrections to the BO approximation up to within linear order in the classical nuclear velocity. We show here that this is  equivalent to a perturbation approach where the small parameter is the electron-nuclear mass ratio. 

The major achievement of such formulation is presented in this paper: electronic non-adiabatic effects appear as a \textit{position-dependent mass} correction to the bare nuclear mass, up to within linear order in the perturbation. From a fundamental perspective, we prove that it is possible to recover an adiabatic-like structure of the Hamiltonian governing the dynamics of the heavy degrees of freedom, with a kinetic energy contribution and a separate potential energy term. Since the mass correction can be fully identified with the electronic mass, totally missing in the BO approximation, we propose a theory able to restore a fundamental property, often overlooked, of the dynamical problem: the translational invariance of an isolated system with its physical mass, i.e. nuclear \textit{and} electronic. If in the BO approximation the nuclear masses are made position-dependent in the way proposed in this paper, the center of mass can be separated from rotations and internal vibrations and evolves as a free particle with mass equal to the total mass of the system (expected from the Galilean invariance of the problem~\cite{Goldhaber_PRA2005}). This property enables us to apply the perturbation approach \textit{before} moving to the molecular center of mass reference frame, with the formal advantage of a very simple and intuitive theory. From an algorithmic perspective, the corrections to the mass involve only ground-state properties and can be calculated as a response to the nuclear motion, within standard perturbation theory~\cite{Baroni_PRL1987, Baroni_PRB1991, Baroni-2001, Gonze-1995b}. Therefore, we are able to perform numerical studies of molecular systems, easily pushing the applications beyond di- and tri-atomic molecules. The experimental implications are clear: the approach proposed here has the potential to predict and to describe ro-vibrational spectroscopic data for a large class of molecular systems when high accuracy is required.

The paper is organized as follows. First we show how, starting from the exact factorization, non-adiabatic effects are included by constructing a perturbative scheme based on the BO approach. Then, we prove that the vector potential of the theory can be expressed as a position-dependent correction to the bare nuclear mass. In the nuclear Hamiltonian, non-adiabatic effects are taken into account in an adiabatic-like picture, if the nuclear masses are corrected for the electronic contribution. We prove that (i) the position-dependent corrections sum up to the total electronic mass of the complete system and (ii) the Hamiltonian with position-dependent dressed masses is appropriate to compute rotational and vibrational spectra as it is possible to exactly separate the center of mass motion. Results are presented, discussing a model of a hydrogen bond and corrections to the vibrational frequencies of small molecular systems.

\section{Beyond the Born-Oppenheimer approximation}\label{sec: bo from xf}
\subsection{Exact factorization of the electron-nuclear wave function}
The exact factorization of the electron-nuclear wave function has been presented~\cite{Gross_PRL2010} and discussed~\cite{Gross_PRL2013, Agostini_ADP2015} in previous work. Therefore, we only introduce here the basic formalism and we refer to the above references for a detailed presentation.

A system of interacting particles, which will be taken as electrons of mass $m_e$ and  nuclei of masses $M_\nu$, is described by the Hamiltonian $\hat H = \hat T_n + \hat H_{BO}$, with $\hat T_n$ the nuclear kinetic energy and $\hat H_{BO}$ the standard BO Hamiltonian. The evolution of the electron-nuclear wave function $\Psi(\br,\bR,t)$, in the absence of an external time-dependent field, is described by the time-dependent Schr\"odinger equation (TDSE) $\hat H \Psi = i\hbar\partial_t\Psi$. The symbols $\br,\bR$ collectively indicate the Cartesian coordinates of $N_{el}$ electrons and $N_n$ nuclei, respectively, in a fix laboratory frame. When the exact factorization is employed, the solution of the TDSE is written as the product $\Psi(\br,\bR,t)=\Phi_{\bR}(\br,t)\chi(\bR,t)$, where $\chi(\bR,t)$ is the nuclear wave function and $\Phi_{\bR}(\br,t)$ is an electronic factor parametrically depending on the nuclear configuration $\bR$. $\Phi_{\bR}(\br,t)$ satisfies the partial normalization condition $\int d\br |\Phi_\bR(\br,t)|^2=1$ $\forall\,\bR,t$, which makes the factorization unique up to a gauge transformation. Starting from the TDSE for $\Psi(\br,\bR,t)$, Frenkel's action principle~\cite{frenkel, Alonso_JCP2013, Gross_JCP2013} and the partial normalization condition yield the evolution equations for $\Phi_{\bR}(\br,t)$ and $\chi(\bR,t)$,
\begin{align}\label{eqn: exact equations}
\left[\hat H_{el}-\epsilon(\bR,t)\right]\Phi_\bR=i\hbar\partial_t\Phi_\bR\quad\mbox{and}\quad\hat H_n\chi=i\hbar\partial_t\chi.
\end{align}
Here, the electronic and nuclear Hamiltonians are $\hat H_{el}=\hat H_{BO}+\hat U_{en}[\Phi_\bR,\chi]$ and $\hat H_n=\sum_\nu[-i\hbar\nabla_\nu+\bA_\nu(\bR,t)]^2/(2M_\nu)+\epsilon(\bR,t)$, respectively. The index $\nu$ is used to label the nuclei. The electron-nuclear coupling operator (ENCO),
\begin{align}\label{eqn: enco}
\hat U_{en}\left[\Phi_\bR,\chi\right]=\sum_{\nu=1}^{N_n}\frac{1}{M_\nu}&\left[\frac{\left[-i\hbar\nabla_\nu-\bA_\nu(\bR,t)\right]^2}{2}+\right.\\
&\left.\left(\frac{-i\hbar\nabla_\nu\chi(\bR,t)}{\chi(\bR,t)}+\bA_\nu(\bR,t)\right)\left(-i\hbar\nabla_\nu-\bA_\nu(\bR,t)\right)\right],\nonumber
\end{align}
the time-dependent vector potential (TDVP),
\begin{align}\label{eqn: vector potential}
\bA_\nu(\bR,t) = \left\langle \Phi_\bR(t)\right| \left.-i\hbar\nabla_\nu\Phi_\bR(t)\right\rangle_\br,
\end{align}
and the time-dependent potential energy surface (TDPES),
\begin{align}\label{eqn: tdpes}
\epsilon(\bR,t) = \left\langle\Phi_\bR(t)\right|\hat H_{BO}+\hat U_{en}-i\hbar\partial_t\left|\Phi_\bR(t)\right\rangle_\br,
\end{align}
mediate the \textit{exact} coupling between the two subsystems, thus they include all effects beyond BO. The symbol $\langle\dots\rangle_\br$ indicates integration over the electronic coordinates. The TDVP and TDPES transform~\cite{Gross_PRL2010} as standard gauge potentials when the electronic and nuclear wave functions transform with a phase $\theta(\bR,t)$. The gauge, the only freedom in the definition of the electronic and nuclear wave functions, will be fixed below.

\subsection{Large nuclear mass limit}
Starting from the XF described above, we now consider the limit of large nuclear masses.
The ENCO is inversely proportional to the nuclear masses $M_\nu$, then the BO limit~\cite{Tully_TCA2000} corresponds to the solution of Eqs.~(\ref{eqn: exact equations}) setting the ENCO to zero~\cite{Scherrer_JCP2015}. Formally, however, approaching this limit of large but finite nuclear masses depends on the physical situation considered~\cite{Hagedorn_AM1986}. In the time-dependent case, keeping fixed the kinetic energy, it has been shown~\cite{Hagedorn_AM1986} that the BO limit is recovered asymptotically in terms of a small expansion parameter $\mu^4$ used to scale the nuclear mass, $M\rightarrow M^{(\mu)}\equiv{M/ \mu^4}$. Making $\mu$ approach zero corresponds to the ratio of the nuclear mass over the electron mass $M^{(\mu)}/m_e$ going to infinity. This scaling factor will be used only to estimate perturbatively the order of the terms in the electronic equation, and will be set equal to unity to recover the values of the physical masses. The nuclear mass being made larger, the nuclear dynamics is slower such that time variable must then be scaled as well, by a factor $\mu^2$, i.e. $t\rightarrow {t / \mu^2}$~\cite{Hagedorn_AM1986}, increasing the separation of time-scales between the light and heavy particles. Similarly, following a simple scaling argument, the nuclear momentum behaves as $\mu^{-2}$ in the semi-classical limit (see Appendix~\ref{app: app 1}). Then, the ENCO from Eq.~(\ref{eqn: enco}) scales with $\mu^4$ as
\begin{align}\label{eqn: encomu}
\hat U_{en,\mu}\left[\Phi_\bR,\chi\right]=\sum_{\nu=1}^{N_n}&\bigg[\frac{\mu^4}{M_\nu}\frac{\left[-i\hbar\nabla_\nu-\bA_\nu(\bR,t)\right]^2}{2}+\\
&\frac{\mu^2}{M_\nu}\bigg(\boldsymbol{\lambda}_\nu(\bR,t)+\mu^2\bA_\nu(\bR,t)\bigg)\Big(-i\hbar\nabla_\nu-\bA_\nu(\bR,t)\Big)\bigg].\nonumber
\end{align}
where $ \boldsymbol{\lambda}_\nu(\bR,t)=\mu^2\frac{-i\hbar\nabla_\nu\chi(\bR,t)}{\chi(\bR,t)}$. $\boldsymbol\lambda_\nu(\bR,t)$ tends towards a quantity independent of $\mu$ in the limit of small $\mu$, since $-i\hbar\nabla_\nu\chi/\chi$ is related to the nuclear momentum~\cite{Scherrer_JCP2015, Agostini_ADP2015} and thus scales as $\mu^{-2}$.

Using the definition in Eq.~(\ref{eqn: tdpes}), we define the scaled TDPES, 
\begin{align}\label{eqn: tdpes mu}
\epsilon_\mu(\bR,t) =& \left\langle\Phi_\bR(t)\right|\hat H_{BO}\left|\Phi_\bR(t)\right\rangle_\br +\mu^2\left\langle\Phi_\bR(t)\right|-i\hbar\partial_t\left|\Phi_\bR(t)\right\rangle_\br\nonumber \\
&+\mu^4\sum_{\nu=1}^{N_n}\frac{1}{2M_\nu}\left\langle\Phi_\bR(t)\right|\left[-i\hbar\nabla_\nu-\bA_\nu(\bR,t)\right]^2\left|\Phi_\bR(t)\right\rangle_\br,
\end{align}
noting the second term in Eq.~(\ref{eqn: encomu}) does not contribute (by construction) to the TDPES. 

\subsection{Perturbative expansion}

The electronic equation thus obtained, 
\begin{align}
&\left[\hat H_{BO}+\hat U_{en,\mu}\left[\Phi_\bR,\chi\right]-\epsilon_\mu(\bR,t)\right] \Phi_\bR=i\hbar\mu^2\partial_t\Phi_\bR,\label{eqn: perturbative electronic eqn}
\end{align}
can be solved perturbatively in powers of $\mu^4$, with its solution of the form $\Phi_\bR(\br,t)=\Phi_\bR^{(0)}(\br,t)+\mu^2\Phi_\bR^{(1)}(\br,t)+\ldots$~\cite{BO, Hagedorn_PSPM2007}. 

The time dependence appears only at order $\mu^2$, as it is clear from Eqs.~(\ref{eqn: tdpes mu}) and~(\ref{eqn: perturbative electronic eqn}). 
Therefore the time dependence of $\Phi_\bR^{(0)}(\br,t)=\varphi_\bR^{(0)}(\br)$ can be dropped out and it satisfies the zeroth order equation 
\begin{align}\label{eqn: BO electronic eqn}
\left[\hat H_{BO}-\epsilon^{(0)}(\bR)\right]\varphi_\bR^{(0)} = 0,
\end{align}
with $\epsilon^{(0)}(\bR)$ the first term on the right-hand-side of Eq.~(\ref{eqn: tdpes mu}). Here, $\varphi_\bR^{(0)}(\br)$ is an eigenstate of the BO Hamiltonian with eigenvalue $\epsilon^{(0)}(\bR)=\epsilon^{(0)}_{BO}(\bR)$, chosen to be the ground state. 

At the zeroth order: (i) the TDVP identically vanishes, $\bA_\nu^{(0)}(\bR,t)=0$, as in the absence of a magnetic field $\varphi_\bR^{(0)}(\br)$ can be taken real; (ii) the evolution of the nuclear wave function is determined by the usual BO equation; (iii) the electronic wave function is used to fix the gauge freedom at all orders, by imposing $\langle \varphi_\bR^{(0)}\vert \Phi_\bR(t)\rangle \in \mathbb{R}$.

The electronic equation at the next order yields
\begin{align}
\left[\hat H_{BO}-\epsilon_{BO}^{(0)}(\bR)\right]\Phi_\bR^{(1)} = i\sum_{\nu=1}^{N_n}\boldsymbol{\lambda}_\nu'(\bR,t)\cdot \left(\hbar\nabla_\nu\varphi_{\bR}^{(0)}\right),\label{eqn: first order a)}
\end{align}
where $\boldsymbol{\lambda}_\nu'(\bR,t)=[\boldsymbol{\lambda}_\nu(\bR,t) + \mu^2\bA_\nu(\bR,t)]/M_\nu$ from Eq.~(\ref{eqn: enco}). We neglected the TDVP from the term in parenthesis since $\bA_\nu(\bR,t)$ is $\mathcal O(\mu^2)$. Furthermore, $\boldsymbol{\lambda}'_\nu$ contains a term $\mathcal O(\mu^2)$, which will be analyzed below along with the TDVP. Appendix~\ref{app: app 2} presents the connection between Eq.~(\ref{eqn: first order a)}) and the nuclear velocity perturbation theory, thus providing a numerical scheme~\cite{Scherrer_JCP2015} to compute $\Phi_\bR^{(1)}(\br,t)$ within perturbation theory~\cite{Scherrer_JCTC2013}. 

The electronic wave function up to within $\mathcal O(\mu^2)$ is
\begin{align}\label{eqn: Phi^(1)}
\Phi_\bR(\br,t)=\varphi_{\bR}^{(0)}(\br)+\mu^2 i\sum_{\nu=1}^{N_n}\boldsymbol{\lambda}_\nu'(\bR,t)\cdot\boldsymbol{\varphi}_{\bR,\nu}^{(1)}(\br),
\end{align} 
where $\boldsymbol{\varphi}_{\bR,\nu}^{(1)}(\br)$ is implicitly defined by Eq.~(\ref{eqn: first order a)}). Eq.~(\ref{eqn: Phi^(1)}) is valid also as initial condition, i.e.~the correction is included if at the initial time the nuclear velocity (the classical limit of $\boldsymbol{\lambda}_\nu'(\bR,t)$) is non-zero~\cite{Rigolin08}.

$\Phi_\bR(\br,t)$ is complex and can thus sustain an electronic current density~\cite{Schild_JPCA2016, Requist_PRA2010} induced by the nuclear motion. The crucial point is that this current influences the nuclear motion through the TDVP.

\subsection{Expression of the time-dependent vector potential}
The TDVP becomes non-zero when inserting Eq.~(\ref{eqn: Phi^(1)}) in Eq.~(\ref{eqn: vector potential}). As described in Appendix~\ref{app: app 3}, Eq.~(\ref{eqn: vector potential}) yields $\ul A(\bR,t)=-\mu^2\dul{{\mathcal A}}(\bR)\ul{{\lambda}}'(\bR,t)$, with
\begin{align}
\dul{{\mathcal A}}(\bR)=2\left\langle\ul{\varphi}_\bR^{(1)}\right|\hat H_{BO}-\epsilon_{BO}^{(0)}(\bR)\left|\ul{\varphi}_\bR^{(1)}\right\rangle_\br.\label{eqn: A-matrix}
\end{align}
The singly underlined symbols $\ul A(\bR,t)$, $\ul{{\lambda}}'(\bR,t)$ and $\ul{\varphi}_\bR^{(1)}$ indicate $(3N_n)$-dimensional vectors, whereas $\dul{{\mathcal A}}(\bR)$ is $(3N_n\times 3N_n)$-dimensional matrix. 

Since $\ul{{\lambda}}'(\bR,t)$ depends on $\ul A(\bR,t)$, we find $\ul A(\bR,t)$ self-consistently, which amounts to include an infinite number of terms of order $\mu^{2n}$. Recalling $\ul{\lambda}=\mu^2 \left[-i\hbar\ul\nabla\chi/\chi\right]$, the TDVP becomes
\begin{align}
\ul{ A}=-\mu^2\dul{{\mathcal A}}\,\dul{{\mathcal M}}^{-1}\ul{{\lambda}},\,\,\mbox{with}\,\,\dul{{\mathcal M}}(\bR)=\dul{M}+\mu^{4}\dul{{\mathcal A}}(\bR).\label{eqn: definition of the mass matrix}
\end{align}
Here, $\dul{M}\equiv M_\nu \delta_{\nu i, \nu^\prime j}$ is the $(3N_n\times 3 N_n)$ diagonal mass matrix. If $\mu^4=1$, expressions where the physical masses appear are recovered. From Eq.~(\ref{eqn: A-matrix}) it is  evident that $\dul{{\mathcal A}}(\bR)$ is a purely electronic quantity, which affects the nuclear momentum through the TDVP. Such correction, however, also appears in the nuclear evolution equation~(\ref{eqn: exact equations}). 

\subsection{Nuclear time-dependent Schr\"odinger equation}
Using the expressions of the TDVP and of the TDPES, as described in Appendix~\ref{app: app 4}, we get an important result: the nuclear TDSE, in matrix form, becomes
\begin{align}\label{eqn: nuclear eqn with position-dependent mass}
\left[\frac{1}{2}\left(-i\hbar\ul\nabla\right)^T\dul{\mathcal M}^{-1}(\bR)\left(-i\hbar\ul\nabla\right)+E(\bR)\right]\chi=i\hbar\partial_t\chi,
\end{align}
where the superscript $T$ indicates the transpose vector and 
\begin{align}
E(\bR)=\epsilon_{BO}^{(0)}(\bR)+\sum_{\nu=1}^{N_n}\frac{\hbar^2}{2M_\nu}\left\langle \nabla_\nu\varphi_\bR^{(0)}\right|\left.\nabla_\nu\varphi_\bR^{(0)}\right\rangle_\br.
\end{align}
The second term is the diagonal BO correction (DBOC). The kinetic energy term now involves dressed nuclear masses.
It is important to notice that such canonical form of the nuclear TDSE arises from the self-consistent solution for $\ul{A}$.

The corresponding classical Hamiltonian~\cite{Goldhaber_PRA2005} is simply $H_n=\ul P^T\dul{\mathcal M}^{-1}(\bR) \ul P/2 + E(\bR)$, with nuclear velocity $\dot{\ul R}=\dul{\mathcal M}^{-1}(\bR)\ul P$. This Hamiltonian contains both the nuclear and electronic contributions to the kinetic energy, in the forms $\dot{\ul{R}}^T\dul M\,\dot{\ul{R}}/2$ and $\dot{\ul{R}}^T\dul{\mathcal A}(\bR)\dot{\ul{R}}/2$, respectively.

The key result of the paper is encoded in Eq.~(\ref{eqn: nuclear eqn with position-dependent mass}), where $\dul{{\mathcal M}}(\bR)=\dul{M}+\dul{{\mathcal A}}(\bR)$ since we have taken $\mu^4=1$. Even in the presence of (weak) non-adiabatic effects, the dynamical problem can be expressed in terms of nuclei moving on a single, static, potential energy surface -- the electronic ground state (plus DBOC) -- with masses that are corrected by the presence of the electrons. We have shown how, in a very simple and intuitive way, the electrons are carried along by the nuclei: $\dul{\mathcal A}(\bR)$, the \Am, is a position-dependent mass that dresses the bare nuclear masses $\dul{M}$. The \Am~is a purely electronic quantity, obtained by considering the lowest order corrections $\mathcal O(\mu^2)$ to the BO electronic wave function, and appears both in the definition of the TDVP and in the nuclear Hamiltonian. The \Am~is the new and fundamental quantity introduced in this study, for which we are able to provide a rigorous derivation, in the context of the exact factorization, an intuitive interpretation, in terms of electronic mass carried along by the motion of the nuclei, and an efficient computation scheme, based on perturbation theory~\cite{Scherrer_JCP2015}.

\section{Properties of the dressed position-dependent mass}\label{sec: properties}
When Cartesian coordinates are employed as done here, the \Am~has the property of yielding the total electronic mass of the system when summed up over all nuclei,
\begin{align}\label{eqn: total electronic mass}
\sum_{\nu,\nu'=1}^{N_n}\mathcal A_{\nu\nu'}^{ij} (\bR)= mN_{el}\delta_{ij} \quad\forall\,\bR,
\end{align}
supporting its interpretation as a correction term to the nuclear mass (indices $\nu$ and $\nu'$ run over the nuclei, $i$ and $j$ over the three spatial dimensions). It should also be noticed that the \Am~is positive-definite in a ground-state dynamics~\cite{Goldhaber_PRA2005}. The proof of Eq.~(\ref{eqn: total electronic mass}) uses the property of the BO electronic wave function of being invariant under a translation of the reference system~\cite{Rauk_CP1993, Goldhaber_PRA2005}, and Eq.~(\ref{eqn: first order a)}) (see Appendix~\ref{app: app 5}). This leads to
\begin{align}\label{eqn: sum rule}
\sum_{\nu,\nu'=1}^{N_n} \left[\dul{{\mathcal A}}(\bR)\right]_{\nu\nu'} =\frac{m}{e}\sum_{\nu=1}^{N_n}\left[\underline{\underline{{\mathcal P}}}(\bR)\right]_{\nu} = mN_{el}\dul{{\mathcal I}}^{(3)},
\end{align}
where $[\dul{{\mathcal A}}(\bR)]_{\nu\nu'}$ and $[\underline{\underline{{\mathcal P}}}(\bR)]_{\nu}$ are $(3\times3)$ matrices (in Cartesian components) and $\dul{{\mathcal I}}^{(3)}$ is the identity matrix. $[\underline{\underline{{\mathcal P}}}(\bR)]_{\nu}=\nabla_{\nu}\langle \hat{\boldsymbol\mu}^{(el)}(\bR)\rangle_{BO}$ is the electronic contribution to the atomic polar tensor, defined as the variation with respect to nuclear positions of the electronic dipole moment (here averaged over the BO state)~\cite{Person1974}. The second equality in Eq.~(\ref{eqn: sum rule}) is obtained using the known property of the atomic polar tensor of yielding the total electronic charge of the system when summed over all nuclei~\cite{Rauk_CP1993,Stephens1990}.

It is common to separate the center of mass (CoM) motion before introducing the BO approximation. Within the molecular frame, the procedure presented here can be straightforwardly applied, by choosing coordinates in which the kinetic energy operator is the sum of two separated terms, i.e. nuclear and electronic. Using the above sum rule, Eq.~(\ref{eqn: sum rule}), it is instead possible to separate of the CoM motion \textit{a posteriori} and recover in that case the full mass of the system.

Starting from the Cartesian coordinates, we make the following change of coordinates
\begin{align}
\begin{array}{ccl}
\bR_1' &=& M_{tot}^{-1}\left(\sum_{\nu=1}^{N_n}M_\nu\bR_\nu + m_e\sum_{k=1}^{N_{el}}\left\langle\hat{\mathbf r}_k\right\rangle_{BO}\right) \label{eqn: CoM}\\
\bR_\nu' &=& \bR_\nu-\bR_1 \quad\mbox{with }\nu\geq2,
\end{array}
\end{align}
with $M_{tot}=\sum_{\nu}M_\nu+m_eN_{el}$. From the sum rule~(\ref{eqn: sum rule}), the nuclear Hamiltonian of Eq.~(\ref{eqn: nuclear eqn with position-dependent mass}) becomes
\begin{align}\label{eqn: Hn with M(R) in internal coordinates}
\hat H_n = \frac{\hat P_{\textrm{CoM}}^2}{2M_{tot}} +  \frac{1}{2}\left(-i\hbar\ul{\nabla}'\right)^T\dul{\mathcal M}'^{-1}\left(-i\hbar\ul{\nabla}'\right) +E'.
\end{align}
$\hat P_{\textrm{CoM}}$ is the momentum (operator) associated to the center of mass (CoM) coordinate in Eq.~(\ref{eqn: CoM}), thus the first term accounts for the motion of the CoM as a free particle. The mass associated to the CoM is, correctly, the total mass of the system, i.e. nuclei and electrons, rather than the nuclear mass only, as in the BO approximation. The following terms in Eq.~(\ref{eqn: Hn with M(R) in internal coordinates}) are the kinetic and potential energies corresponding to the internal, rotational and vibrational, degrees of freedom (see Appendix~\ref{app: app 6} for a detailed derivation).

\section{Applications}\label{sec: results}
The formalism introduced above is employed to construct a numerical procedure that is (i) fundamentally adiabatic, namely only a single (static) potential energy surface is explicitly involved, but (ii) able to account for electronic effects beyond BO via the position-dependent corrections to the bare nuclear masses. The key quantity in the examples reported below is the nuclear Hamiltonian of Eq.~(\ref{eqn: nuclear eqn with position-dependent mass}): quantum-mechanically, it will be used to compute the spectrum of a model of a proton involved in a one-dimensional hydrogen bond~\cite{Borgis_CPL2006}; interpreted classically in the same model system, it will be employed as the generator of the classical evolution of the oxygen atoms in the presence of a quantum proton. Transforming to internal coordinates and within the harmonic approximation, position-dependent corrections are included in the calculation of the vibrational spectra of H$_2$, H$_2$O, NH$_3$ and H$_3$O$^+$. Numerical details are given in Appendix~\ref{app: app 7}.

\subsection{Proton transfer}\label{sec: results 1}
\begin{figure}[h]
\begin{center}
\centerline{\includegraphics[width=.4\textwidth]{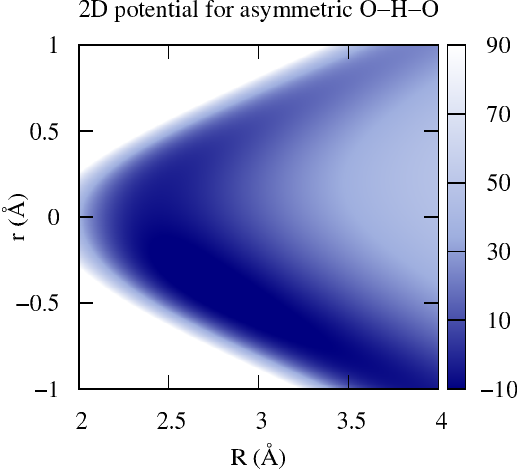}}
\caption{\label{fig: panel a}Potential of the hydrogen bond model as a function of the O$-$O distance ($R$) and of the proton position ($r$).}
\end{center}
\end{figure}

\begin{figure}[h]
\begin{center}
\centerline{\includegraphics[width=.4\textwidth]{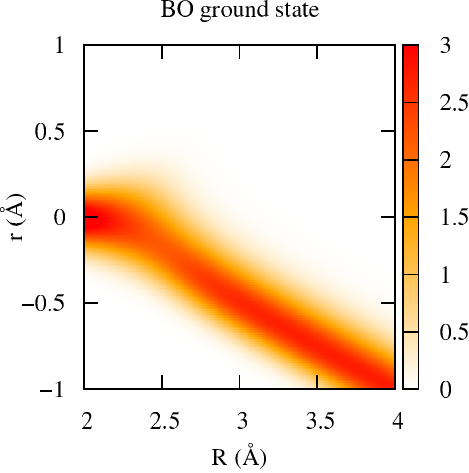}}
\caption{\label{fig: panel b} Proton density corresponding to the BO ground state.}
\end{center}
\end{figure}

\begin{figure}[h]
\begin{center}
\centerline{\includegraphics[width=.4\textwidth]{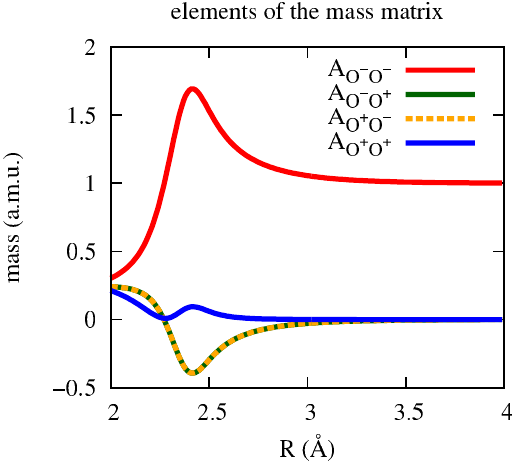}}
\caption{\label{fig: panel c} Elements of the \Am~as functions of $R$.}
\end{center}
\end{figure}

\begin{figure}[h]
\begin{center}
\centerline{\includegraphics[width=.4\textwidth]{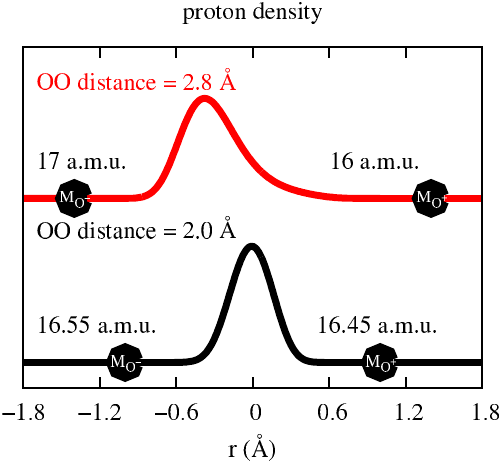}}
\caption{\label{fig: panel d} Proton density at two values of the O$-$O distance (2.0~\AA~black and 2.8~\AA~red), where the masses of the oxygens (sum of columns of the matrix $\dul{\mathcal M}$, see text) $\mathcal M_{\mathrm O^+}$ and $\mathcal M_{\mathrm O^-}$ indicate the \Am~effect.}
\end{center}
\end{figure}

As a first application, we consider a model of a proton involved in a one-dimensional hydrogen bond O$-$H$-$O~\cite{Borgis_CPL2006}, in which non-adiabatic effects are known to be important~\cite{Kapral05}. The light particle is the proton, assumed to be in its vibrational ground state. The mass ratio with the heavy particles, the two oxygens, is much larger than the electron-nuclear mass ratio, thus suggesting possible deviations from the BO approximation. We use an asymmetric potential mimicking a strong hydrogen bond (as shown in Fig.~\ref{fig: panel a}): the proton is bonded to the oxygen atom O$^-$ at large distances (we denote O$^-$ the oxygen atom that is located on the left and O$^+$ the one on the right) whereas at short distances it is shared by the two oxygen atoms and is localized around the center of the O$-$O bond. 
The proton density corresponding to the ground state is shown in Fig.~\ref{fig: panel b}. 
At large distances we expect the effective mass of  O$^-$ to be close to 17~a.m.u. as it carries along the proton. This is clear in Fig.~\ref{fig: panel c}, where it is shown that the element $\mathcal A_{\mathrm O^-\mathrm O^-}(R)$ of the \Am~tends to a constant (equal to 1~a.m.u., the mass of the proton) at $R>3$~\AA, whereas all other components are zero, as expected from the sum rule of Eq.~(\ref{eqn: sum rule}). 
We show this schematically in Fig.~\ref{fig: panel d} where we plot the proton density along the O$-$O bond. We also report an estimate of the amount of electronic mass associated to each oxygen, as the sum over the columns of the \Am, e.g. $\mathcal M_{\mathrm O^-}(R)=M_{\mathrm O^-}+[\mathcal A_{\mathrm O^-\mathrm O^-}(R)+\mathcal A_{\mathrm O^+\mathrm O^-}(R)]$.
At short distances instead the proton is shared by the oxygens: the elements of the \Am~are non-zero, but the $\mathrm O^-$ diagonal contribution remains dominant. Notice that it is not surprising that the off-diagonal elements of the \Am~are negative, as only two conditions are physically relevant: the diagonal elements must be non-negative, in a ground-state dynamics, and the sum of the elements must yield the electronic mass, in a translationally invariant system. As seen in Fig.~\ref{fig: panel d}, the two oxygens have then similar masses at very short distances.

\begin{figure}[h]
\begin{center}
\centerline{\includegraphics[width=.4\textwidth]{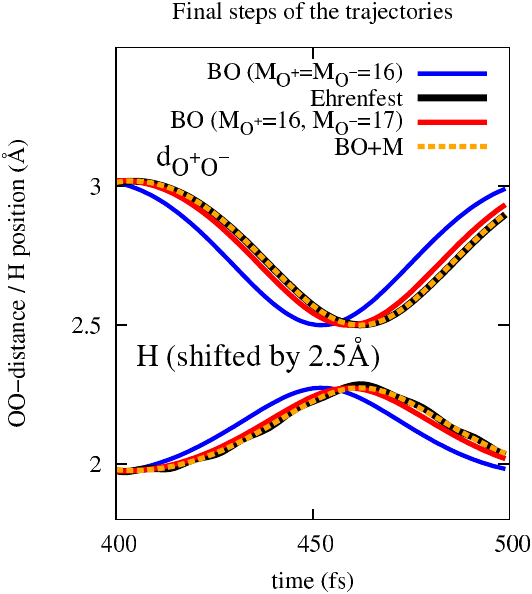}}
\caption{\label{fig: panel e} Distance of the oxygens and position of the proton during the final steps of the dynamics: BO approximation with $M_{\textrm{O}^+}=M_{\textrm{O}^-}=16$~a.m.u. (blue), BO approximation adding the proton mass $M_{\textrm H}=1$~a.m.u. to the the mass of $M_{\textrm{O}^-}$ (red), the BO approximation corrected by the position-dependent dressed mass (orange), Ehrenfest dynamics (black).}
\end{center}
\end{figure}

\begin{figure}[h]
\begin{center}
\centerline{\includegraphics[width=.4\textwidth]{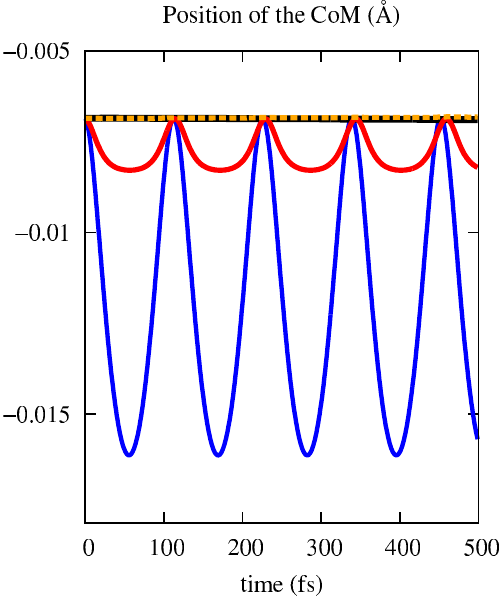}}
\caption{\label{fig: panel f} Position of the CoM.}
\end{center}
\end{figure}

\begin{figure}[h]
\begin{center}
\centerline{\includegraphics[width=.4\textwidth]{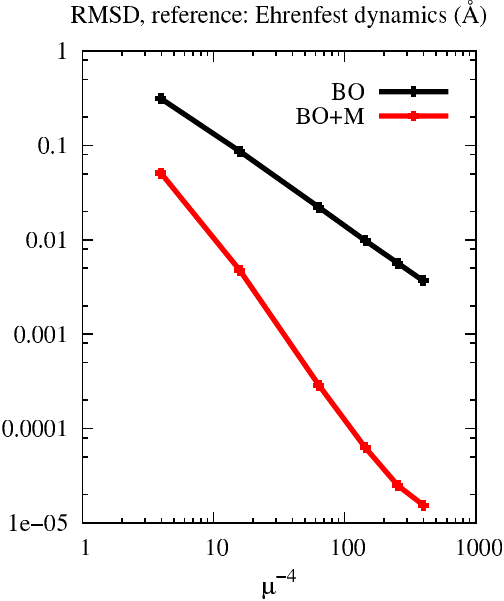}}
\caption{\label{fig: panel g} RMSD between Ehrenfest results and the BO approximation (black) or the BO approximation corrected by the position-dependent dressed mass (red). The results are shown as functions of the inverse mass ratio $\mu^{-4}$.}
\end{center}
\end{figure}

Fig.~\ref{fig: panel e} shows the classical trajectories of the two oxygen atoms starting from a compressed O$-$O distance and zero velocity. 
Calculations have been performed both in the standard adiabatic approximation (BO) and with position-dependent corrections to the oxygen masses (BO$+$M). The two sets of calculations are compared with Ehrenfest dynamics, where non-adiabatic effects are included explicitly. Ehrenfest-type simulations, being explicitly non-adiabatic, require calculations of electronic excited-state quantities, limiting not only the size of the accessible systems but also the time-scales. BO$+$M calculations, based on the perturbation to the electronic ground state, are instead easily affordable. The distance of the oxygens is plotted along with the mean position of the proton at the final steps of the dynamics. The masses are M$_{\textrm{O}^+}=$ M$_{\textrm{O}^-}=16$~a.m.u. and M$_{\textrm{H}^+}=1$~a.m.u. In Fig.~\ref{fig: panel f} it is shown that the CoM of the system is perfectly fixed when position-dependent masses are employed, in contrast to the BO approximation. BO dynamics is faster than the Ehrenfest dynamics because the heavy atoms have only the bare nuclear mass. We have tested an \textit{ad hoc} correction to the mass of the oxygen O$^-$, i.e. M$_{\textrm{O}^-}=17$~a.m.u. 
This improves the conservation of the CoM but does not fix it completely. Changing M$_{\textrm{O}^-}$ to 17~a.m.u. improves the result, but only including the position-dependent dressed mass leads to a systematic convergence to the Ehrenfest results. We have further compared the error with respect to Ehrenfest dynamics, of BO and BO+M dynamics, as function of the inverse mass ratio $\mu^{-4}=M_{\mathrm O}/M_{\textrm{H}^+}$. This is shown  in Fig.~\ref{fig: panel g} as the root-mean-square-deviation (RMSD) with respect to the reference Ehrenfest trajectory. 
The position-dependent dressed mass greatly improves the precision of the dynamics even at small values of $\mu^{-4}$ ($=4$ is the smallest value used), and leads to an error four orders of magnitude smaller than BO at large mass ratios.

Compared to Ehrenfest dynamics, the BO+M dynamics is much less computationally expensive, having a similar cost as the BO dynamics itself. Furthermore, the proposed Hamiltonian formulation allows for a full quantum treatment of the nuclear dynamics: it maintains the simplicity of the BO approximation, making feasible calculations of large systems, while at the same time it gains in accuracy.

\begin{figure}[h]
\begin{center}
\centerline{\includegraphics[width=.4\textwidth]{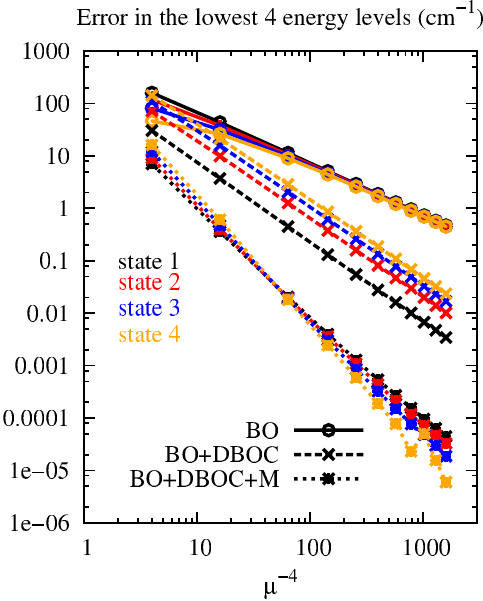}}
\caption{\label{fig: panel h} Error between the four lowest eigenvalues of the full Hamiltonian and BO (solid lines with circles), BO+DBOC (dashed lines with crosses), BO+DBOC+M (dotted lines with squares). The results are shown as functions of the inverse mass ratio $\mu^{-4}$.}
\end{center}
\end{figure}

To illustrate this, we have computed the four lowest eigenstates of the full quantum Hamiltonian at different values of $\mu^{-4}$. The diagonalization of the full Hamiltonian is compared to three approximations: BO, BO+DBOC, BO+DBOC+M (where we also include the position-dependent correction). Fig.~\ref{fig: panel h} shows the error on the eigenvalues (the exact lowest eigenvalue is $-4127.08527$~cm$^{-1}$ at M$_{\textrm{O}^+}=$ M$_{\textrm{O}^-}=16$~a.m.u.). At small $\mu^{-4}$ the BO approximation is expected to fail: the mass corrections allow to gain one order of magnitude in the eigenvalues, even if compared to the case where the DBOC is included. Overall, also in the static situation the mass correction leads to highly accurate results. At a mass ratio $\mu^{-4}=1600$ an accuracy on the eigenvalues of about $10^{-5}$~cm$^{-1}$ is reached whereas it is only $0.5$~cm$^{-1}$ using the BO approximation.

\subsection{Corrections to harmonic frequencies}\label{sec: results 3}
Next, we consider non-adiabatic effects on vibrational frequencies and predict corrections to the harmonic frequencies of small molecules, i.e.~H$_2$, H$_2$O, NH$_3$ and H$_3$O$^+$. The corrected frequencies $\nu+\Delta\nu$ have been computed by diagonalizing the matrix $[\dul{{\mathcal M}}^{-1}(\bR_0) \dul{K}(\bR_0)]$ at the equilibrium geometry $\bR_0$, where $\dul{K}$ is the Hessian computed from the ground state adiabatic potential energy surface. Negative frequency shifts are expected, as shown in Table~\ref{tab: frequencies}: non-adiabatic effects perturbing the ground-state dynamics tend to induce excitations of the light particles, and the energy necessary for the transition is ``removed'' from the heavy particles. Comparison with the literature~\cite{Schwenke_JPCA2001}, when available, shows that the theory is capable to predict accurate non-adiabatic corrections, even if working within the harmonic approximation and with the generalized gradient approximation to density functional theory. What the approach developed in this study contributes to the field is the possibility of easily extending the numerical applications beyond di- and tri-atomic molecules. To prove this, we provide the first predictions to the non-adiabatic corrections of vibrational frequencies of NH$_3$ and H$_3$O$^+$. It can be seen that the shifts of the N-H stretch frequencies of NH$_3$ are larger than those of the O-H stretch frequencies of H$_3$O$^+$, due to the fact that the N-H bonds are less ionic and as a result the mass carried along by the protons are larger in NH$_3$ than in H$_3$O$^+$.
\begin{table}[h]
\begin{tabular}{@{\vrule height 10.5pt depth4pt  width0pt}c|cc|cc|c|c}
%molecule & $\nu$  & $\Delta\nu$ & reference $\nu$ and $\Delta\nu$ \\
\hline
\hline
molecule & H$_2$ & H$_2$~\cite{Schwenke_JPCA2001} & H$_2$O & H$_2$O~\cite{Schwenke_JPCA2001} & NH$_3$ & H$_3$O$^+$ \\
\hline
$\nu$, $\Delta\nu$ & 4343.28, -0.89 & -0.74 & 1594.93, -0.06 & 1597.60, -0.07 & 1016.73, -0.06 & 837.27, -0.03 \\
                              &                          && 3656.19, -0.74 & 3661.00, -0.69 & 1628.30, -0.10 & 1639.25, -0.05 \\
                              &                          && 3757.77, -0.59 & 3758.63, -0.77& 3358.91, -0.82 & 3438.80, -0.36 \\
                              &                          &&                         & & 3471.93, -0.74 & 3522.10, -0.26 \\
\hline
% &  & 1597.60, -0.07~\cite{Schwenke_JPCA2001} \\
%&  &3661.00, -0.69~\cite{Schwenke_JPCA2001} \\
%&  & 3758.63, -0.77~\cite{Schwenke_JPCA2001} \\
\hline
\end{tabular}
\caption{\label{tab: frequencies}Harmonic frequencies $\nu$ (in $cm^{-1}$) and their non-adiabatic corrections $\Delta\nu$. Benchmark values are taken from~\cite{Schwenke_JPCA2001} when indicated.}
\end{table}

\subsection{Methods}
In the static calculations of the O$-$H$-$O model, the eigenvalues of the full Hamiltonian are determined using a Gaussian quadrature method with 20 points for $R$, the O$-$O distance, and 34 for $r$, the position of the proton from the center of the O$-$O bond. In the dynamics we use the three coordinates, i.e. $R_{\mathrm{O}^+}$, $R_{\mathrm{O}^-}$ and $r=r_{\mathrm{H}}$, in order to test the conservation of the position of the CoM. The velocity-Verlet algorithm is used to integrate the classical nuclear equations, with a time-step of 1~fs; the Crank-Nicolson~\cite{Nicolson_PCPS1947} algorithm for the proton (quantum) equation in Ehrenfest dynamics, with a time-step of 10$^{-4}$~fs; the Euler algorithm with time-step 0.0625~fs for BO+M calculations, where the force depends on the velocity. For the vibrational spectra, the \Am~has been computed using density functional perturbation theory~\cite{Scherrer_JCP2015,Scherrer_JCTC2013,Putrino-2000} and it has been checked that the sum rule of Eq.~(\ref{eqn: sum rule}) is satisfied. The numerical scheme has been implemented in the electronic structure package CPMD~\cite{CPMD}. Calculations have been performed using Troullier-Martins~\cite{Troullier-1991} pseudo-potentials in the Becke-Lee-Yang-Parr~\cite{Becke1988, Lee1988} (BLYP) approximation of the exchange-correlation kernel. The equilibrium molecular geometry is determined at the BLYP level, employing the aug-cc-pVTZ basis set~\cite{Kendall1992} in the Gaussian electronic structure program~\cite{g09}.

\section{Conclusions}\label{sec: conclusions}
This work provides a rigorous theory to include the effect of electronic motion on nuclear dynamics in molecules within the adiabatic framework. Nuclear masses are dressed by position-dependent corrections that are purely electronic quantities and a consequence of the fact that electrons do not follow rigidly the motion of the nuclei. Various applications are discussed, yielding in all cases striking agreement with the benchmarks, either exact or highly accurate quantum-mechanical calculations. The idea of including perturbatively non-adiabatic electronic effects on the nuclear motion has been previously introduced~\cite{Moss_MP1977, Tennyson_JCP1999, Wolniewicz_JCP1995, Nafie_JCP1983, Schild_JPCA2016, Scherrer_JCP2015}, mainly as a tool to resolve some of the issues encountered in the context of theoretical vibrational spectroscopy when working in the BO approximation. Similarly, the idea of accounting for corrections to the nuclear masses has been proposed~\cite{Moss_MP1977, Tennyson_JCP1999, Kutzelnigg_MP2007, Goldhaber_PRA2005} to cure some fundamental inconsistencies of the BO treatment. The novelty of the present study is thus to be found in the overall picture that our work conveys: the theory is developed based on a rigorous starting point, the exact factorization of the molecular wave function; the perturbation treatment is justified in terms of the electron-nuclear mass ratio, as in the seminal paper of Born and Oppenheimer; the algebraic procedure is very simple, easily allowing for applications not restricted to di- and tri-atomic molecules; the proposed numerical scheme requires standard electronic structure calculations to determine the mass corrections, as the expression of such corrections are explicitly given in terms of electronic properties. We expect that the theory will be able to provide solid information to predict and interpret highly accurate spectroscopy experiments on a large class of molecular systems.

Conceptually, we have resolved a well-known~\cite{Subotnik_JPCL2012} fundamental inconsistency of the BO approximation. In a translationally invariant problem, the CoM moves as a free particle with mass that equals the total mass of the systems, i.e. nuclei and electrons, not only the nuclear mass. This feature is naturally built in the theory and corrects for a deficiency of the BO approximation, providing exactly the missing mass of the electrons. From a more practical point of view, our approach is very general and can be applied whenever a ``factorization'' of the underlying physical problem is possible, e.g.~in the case of proton and oxygen atoms or in the case of electrons and nuclei.

Further applications are indeed envisaged, since the perturbative incorporation of non-adiabatic effects greatly reduces the complexity of the fully coupled problem. For instance, the approximations can be applied to nuclear wave packet methods for the calculation of highly accurate vibrational spectra beyond the BO approximation.
The position dependent mass is also shown to be related to the ionicity of the bonds and may serve as a proxy to access electronic properties.

%\addcontentsline{toc}{section}{References}
%\bibliography{./biblio,./factorization}

\appendix

\section{The adiabatic limit of the exact factorization}\label{app: app 1}
We will argue in this section that (i) the correct scaling of the time variable is $\mu^2$, when the parameter $\mu^4$ is used to scale the nuclear mass $M_\nu$, and (ii) the term in the electron-nuclear coupling operator of Eq.~(2) containing the nuclear wave function scales as well with $\mu^2$.

Statement (i) is obtained by taking the large mass limit, or small $\mu^4$ limit, as in~\cite{Hagedorn_AM1986}. In this situation, the dynamics of the heavy nuclei becomes semi-classical and our scaling argument will make the nuclear kinetic energy tend towards a constant. In the classical limit, it is easy to see that at different values of $\mu^{4}$ the trajectories of the nuclei can be superimposed if the physical time $s$ is rescaled to a common time $t=\mu^2 s$. At each configuration $\ul R(s)$ along the dynamics, the scaling of the time variable has the effect of yielding a kinetic energy that is a constant of $\mu^4$. In other words, the velocities $\ul V(s)$  scale as $\mu^{-2}$. Notice that this is possible as we do not scale the positions with $\mu^4$, and therefore the potential energy is not affected by the scaling. Using the common rescaled time $t$ to describe the nuclear trajectory, it then becomes possible to make a convergence statement about the nuclear dynamics.

Following Ref.~\cite{Hagedorn_AM1986}, the nuclear wave packet can be considered to be a Gaussian wave packet localized at the position $\ul{R}(t)$ , with momentum $\ul{P}(t)$:
\begin{align}
\chi(\ul R,t)&=
\pi^{-3N_n/4}\mu^{-3N_n/2}(\det\dul{{\sigma}}(t))^{1/4}e^{\left[-\frac{(\ul R-\ul R(t))^T \dul{{\sigma}}(t) (\ul R-\ul R(t))}{2\mu^2}+\frac{i}{\hbar} \left(\frac{\ul{{P}}(t)}{\mu^2}\right)\cdot(\ul R-\ul R(t))\right]},
\end{align}
with $\dul{{\sigma}}(t)$ a $(3N_n\times 3N_n)$ symmetric matrix yielding the spatial extension of the wave packet. From this expression, we see that statement (ii) holds: $-i\hbar\nabla_\nu\chi/\chi$ scales as $\mu^{-2}$ thus $\displaystyle  \boldsymbol\lambda_\nu(\bR,t)=   \mu^2 \frac{-i\hbar\nabla_\nu\chi(\bR,t)}{\chi(\bR,t)}$ tends towards a quantity independent of $\mu$.

\section{Nuclear velocity perturbation theory}\label{app: app 2}
In this section we show the relation between the $\mu^4-$expansion proposed in the paper and the nuclear velocity perturbation theory (NVPT) of Ref.~\cite{Scherrer_JCP2015}. We recall here the definition of $\boldsymbol{\lambda}_\nu'(\bR,t)$, 
\begin{align}\label{eqn: def parameter}
\boldsymbol{\lambda}_\nu'(\bR,t)&=\frac{1}{M_\nu}\left(\mu^2\frac{-i\hbar\nabla_\nu\chi(\bR,t)}{\chi(\bR,t)}+\mu^2\bA_\nu(\bR,t)\right)\\
&=\frac{1}{M_\nu}\big(\boldsymbol\lambda_\nu(\bR,t)+\mu^2\bA_\nu(\bR,t)\big).
\end{align}
In the framework of NVPT we have used $\boldsymbol\lambda_\nu(\bR,t)/M_\nu$ as the perturbation parameter that controls the degree of non-adiabaticity of the problem. The electronic equation~(7) can be written using $\boldsymbol{\lambda}_\nu'(\bR,t)$ as
\begin{align}\label{eqn: el eqn with mu}
\left[\hat H_{BO}-\epsilon_{BO}^{(0)}(\bR)\right]\left(\varphi_\bR^{(0)}(\br)+\mu^2\Phi_\bR^{(1)}(\br,t)\right)=\mu^2\sum_{\nu=1}^{N_n}\boldsymbol{\lambda}_\nu'(\bR,t)\cdot\left[i\hbar\nabla_\nu+{\bf A}_\nu(\bR,t)\right]\varphi_\bR^{(0)}(\br).
\end{align}
Also, as will be proved in Appendix~\ref{app: app 3}, the time-dependent vector potential (TDVP) is itself $\mathcal O(\mu^2)$, thus it will be neglected from the term in square brackets on the right-hand-side. If we solve this equation order by order, Eqs.~(8) and~(9) are easily obtained. In particular, we recall here Eq.~(9) whose solution yields $\Phi_\bR^{(1)}(\br,t)$,
\begin{align}
\left[\hat H_{BO}-\epsilon_{BO}^{(0)}(\bR)\right]\Phi_\bR^{(1)}(\br,t) = i\sum_{\nu=1}^{N_n}\boldsymbol{\lambda}_\nu'(\bR,t)\cdot \left(\hbar\nabla_\nu\varphi_{\bR}^{(0)}(\br)\right).\label{eqn: first order a in app)}
\end{align}
In Ref.~\cite{Scherrer_JCP2015} we started from the electronic Hamiltonian of the form
\begin{align}\label{eqn: perturbed el H in nvpt}
\hat H_{el} = \hat H_{BO} + \sum_{\nu=1}^{N_n}\boldsymbol{\lambda}_\nu'(\bR,t)\cdot \left(-i\hbar\nabla_\nu\right),
\end{align}
and we have solved it perturbatively, using $\hat H_{BO}$ as the unperturbed Hamiltonian. It is clear, as stated above, that $\boldsymbol{\lambda}_\nu'(\bR,t)$ is the small parameter that controls the strength of the perturbation and that $-i\hbar\nabla_\nu$ is the (non-adiabatic) perturbation. We have looked for the eigenstates of $\hat H_{el}$ in the form
\begin{align}
\Phi_\bR(\br,t) = \varphi_\bR^{(0)}(\br) +\sum_{e\neq0}\frac{\left\langle\varphi_\bR^{(e)}\right|-i\hbar\sum_\nu\boldsymbol{\lambda}_\nu'(\bR,t)\cdot\nabla_\nu\left.\varphi_\bR^{(0)}\right\rangle_\br}{\epsilon_{BO}^{(0)}(\bR)-\epsilon_{BO}^{(e)}(\bR)}\varphi_\bR^{(e)}(\br),
\end{align}
as straightforwardly follows from the application of standard time-independent perturbation theory. The first order perturbation to the BO ground state can be written as
\begin{align}\label{eqn: nacvs in phi 1}
i\boldsymbol{\varphi}_\bR^{(1)}(\br) = i\sum_{e\neq0} \frac{\mathbf d_{\nu,e0}(\bR)}{\omega_{e0}(\bR)}\varphi_\bR^{(e)}(\br)
\end{align}
with $\omega_{e0}(\bR) = (\epsilon_{BO}^{(e)}(\bR)-\epsilon_{BO}^{(0)}(\bR))/\hbar$ and $\mathbf d_{\nu,e0}(\bR) = \langle\varphi_\bR^{(e)}|\nabla_\nu\varphi_\bR^{(0)}\rangle_\br$, the non-adiabatic coupling vectors. This leads to a new expression of $\Phi_\bR(\br,t)$,
\begin{align}\label{eqn: perturbed el wf}
\Phi_\bR(\br,t) = \varphi_\bR^{(0)}(\br) + i\sum_{\nu=1}^{N_n} \boldsymbol{\lambda}_\nu'(\bR,t)\cdot\boldsymbol{\varphi}_\bR^{(1)}(\br),
\end{align}
which is exactly Eq.~(10) when setting $\mu^2=1$, to obtain the physical nuclear mass.

In the framework of NVPT, the perturbation parameter has been interpreted classically as the nuclear velocity~\cite{Scherrer_JCP2015, Gross_JCP2014,Agostini_ADP2015}. It is worth mentioning here that, when performing a numerical simulation, such dependence on the nuclear velocity shall be correctly accounted for, also in the preparation of the initial electronic state. When using NVPT to perform the calculations, the electronic evolution is not explicit, in the sense that at each time the electronic wave function is simply reconstructed using ground state properties that are then inserted in Eq.~(\ref{eqn: perturbed el wf}). However, when NVPT results are (or can be) compared with quantum-mechanical fully non-adiabatic results, the initial electronic state cannot be simply prepared in the ground state, unless the initial nuclear velocity is zero. If this is not the case, then the first order contribution in Eq.~(\ref{eqn: perturbed el wf}), proportional to the finite value of the initial nuclear velocity, has to be included in the initial condition. Then NVPT and non-adiabatic results can be directly compared, as the same initial conditions are used in both.

Equating the first order corrections to the BO eigenstate, from the $\mu^4-$ and the $\boldsymbol{\lambda}_\nu'(\bR,t)-$ expansion, yields
\begin{align}
\Phi_\bR^{(1)}(\br,t) = i\sum_{\nu=1}^{N_n} \boldsymbol{\lambda}_\nu'(\bR,t)\cdot\boldsymbol{\varphi}_\bR^{(1)}(\br).
\end{align}
The comparison between the $\mu^4-$expansion and NVPT allows, first of all, to derive an explicit expression of $\boldsymbol{\varphi}_\bR^{(1)}(\br)$ as given in Eq.~(\ref{eqn: nacvs in phi 1}), and, second, to decompose the perturbed state as a sum of independent (linear) responses to the non-adiabatic perturbations, thus leading to
\begin{align}\label{eqn: first order electronic eqn}
\left[\hat H_{BO}-\epsilon_{BO}^{(0)}(\bR)\right]{\varphi}_{\bR,\nu\alpha}^{(1)}(\br) = \hbar\partial_{\nu\alpha}\varphi_{\bR}^{(0)}(\br).
\end{align}
As above, the index $\nu$ is used to label the nuclei and $\alpha$ labels the Cartesian components of the gradient. This equation can now be easily solved by employing density functional perturbation theory as described in Ref.~\cite{Scherrer_JCP2015}.

\section{Analysis of the perturbation parameter}\label{app: app 3}
The TDVP, defined in Eq.~(3), is written using Eq.~(10) as
\begin{align}
\bA_\nu(\bR,t) =\left\langle\varphi_{\bR}^{(0)}+i\mu^2\sum_{\nu'=1}^{N_n}\boldsymbol\lambda_{\nu'}'(\bR,t)\cdot\boldsymbol\varphi_{\bR,\nu'}^{(1)}\right|\left.-i\hbar\nabla_\nu\varphi_{\bR}^{(0)}+\mu^2\hbar\nabla_\nu\sum_{\nu'=1}^{N_n}\boldsymbol\lambda_{\nu'}'(\bR,t)\cdot\boldsymbol\varphi_{\bR,\nu'}^{(1)}\right\rangle_\br.
\end{align}
Up to within the linear order in $\mu^2$ (or more precisely $\mu^2 \boldsymbol\lambda_{\nu'}'(\bR,t)$), this expression is 
\begin{align}
\bA_\nu(\bR,t) = -2\hbar\mu^2\int d\br\sum_{\nu'=1}^{N_n}\left[\boldsymbol\lambda_{\nu'}'(\bR,t)\cdot\boldsymbol\varphi_{\bR,\nu'}^{(1)}(\br)\right]\nabla_\nu\varphi_{\bR}^{(0)}(\br)
\end{align}
where we can use Eq.~(\ref{eqn: first order electronic eqn}) to identify the \Am,
\begin{align}\label{eq:a_mat_calc}
\dul{{\mathcal A}}(\bR)=2\left\langle\ul{\varphi}_\bR^{(1)}\right|\hat H_{BO}-\epsilon_{BO}^{(0)}(\bR)\left|\ul{\varphi}_\bR^{(1)}\right\rangle_\br.
\end{align}
We derive the following expression of the TDVP, namely
\begin{align}
\ul{A}(\bR,t) = -\mu^2\dul{{\mathcal A}}(\bR)\ul{{\lambda}}'(\bR,t).
\end{align}
Once again we keep the term $\mathcal O(\mu^2)$ in $\ul{\lambda}'$, but we will show below how it will be included in the definition of the small parameter $\ul{\lambda}$. $\dul{{\mathcal A}}(\bR)$ is a matrix, thus the double-underlined notation, with ($3N_n\times 3N_n$) elements, whereas $\ul{\varphi}_\bR^{(1)}(\br)$ is a vector with ($3N_n$) components. We have written also the TDVP and the parameter in matrix notation, with $\ul{A}(\bR,t)$ and $\ul{{\lambda}}'(\bR,t)$ $(3N_n)-$dimensional vectors. The elements of the \Am~are
\begin{align}
\mathcal A^{ij}_{\nu'\nu}(\bR)=\left\langle\varphi_{\bR,\nu'i}^{(1)}\right|\hat H_{BO}-\epsilon_{BO}^{(0)}(\bR)\left|\varphi_{\bR,\nu j}^{(1)}\right\rangle_\br,
\end{align}
with $i,j$ labeling the Cartesian components and $\nu',\nu$ the nuclei. When using Eq.~(\ref{eqn: nacvs in phi 1}), the elements of the \Am~can be written in terms of the non-adiabatic coupling vectors and of the BO eigenvalues as
\begin{align}
\mathcal A^{ij}_{\nu'\nu}(\bR)=2\hbar\sum_{e\neq0}\frac{d_{\nu'i,e0}(\bR)d_{\nu j,e0}(\bR)}{\omega_{e0}(\bR)}
\end{align}
from which it follows that the \Am~is symmetric. The \Am~is also positive definite (i.e. for all non-zero real vectors $\ul v$, the relation $ \ul v^T \dul{\mathcal A}\ul v\geq 0 $ holds) with non-negative diagonal elements, i.e.
\begin{align}
\mathcal A^{ii}_{\nu\nu}(\bR)=2\hbar\sum_{e\neq0}\frac{\left|\mathbf d_{\nu i,e0}(\bR)\right|^2}{\omega_{e0}(\bR)}\geq0.
\end{align}
This property is essential for the interpretation of the \Am~as a position-dependent mass. The components of the TDVP can be expressed in terms of the components of the \Am,
\begin{align}\label{eqn: vector potential with A-matrix}
A_{\nu i}(\bR,t) = -\mu^2\sum_{\nu'=1}^{N_n}\sum_{j=x,y,z} \mathcal A^{ij}_{\nu\nu'}(\bR)\lambda_{\nu' j}'(\bR,t).
\end{align}
This expression is used in the definition of the parameter $\lambda_{\nu i}'(\bR,t)$, given in Eq.~(\ref{eqn: def parameter}),
\begin{align}
\lambda_{\nu i}'(\bR,t)=M_\nu^{-1}\lambda_{\nu i}(\bR,t)-\mu^4M_\nu^{-1}\sum_{\nu',j} \mathcal A^{ij}_{\nu\nu'}(\bR)\lambda_{\nu' j}'(\bR,t),\label{eqn: components of l'}
\end{align}
where 
\begin{align}
\lambda_{\nu i}(\bR,t)=\mu^2\frac{-i\hbar\partial_{\nu i}\chi(\bR,t)}{\chi(\bR,t)},
\end{align}
which, we recall, tends towards a quantity independent of $\mu$ if $\mu\rightarrow 0$.

Writing Eq.~(\ref{eqn: components of l'}) in matrix form and solving for $\ul{\lambda}(\bR,t)$ we obtain
\begin{align}\label{eqn: definition of the mass matrix in app}
\ul{\lambda}(\bR,t)=\left[\dul{M}+\mu^4\dul{{\mathcal A}}(\bR)\right]\ul{\lambda}'(\bR,t)=\dul{{\mathcal M}}(\bR)\ul{\lambda}'(\bR,t),
\end{align}
where $\dul{M}$ is a diagonal ($3N_n\times 3N_n$) matrix containing the masses of the nuclei and we have defined a position-dependent mass matrix $\dul{{\mathcal M}}(\bR)$. This equation can be inverted to obtain
\begin{align}\label{eqn: matrix relation between the parameters in app}
\ul{\lambda}'(\bR,t)=\dul{{\mathcal M}}^{-1}(\bR)\ul{\lambda}(\bR,t),
\end{align}
yielding the TDVP in the form given in Eq.~(12)
\begin{align}\label{eqn: tdvp}
\ul{A}(\bR,t) = -\dul{{\mathcal A}}(\bR)\dul{{\mathcal M}}^{-1}(\bR)\ul{\lambda}(\bR,t)
\end{align}
with $\mu^4=1$, where only $\ul\lambda$ appears.

Eq.~(\ref{eqn: vector potential with A-matrix}) shows that the TDVP is at least first order in the perturbation parameter and this is the reason why it is not considered in the definition of the perturbed electronic Hamiltonian in Eq.~(\ref{eqn: perturbed el H in nvpt}). Due to the explicit dependence of $\bA_\nu(\bR,t)$ on $\boldsymbol{\lambda}_\nu'(\bR,t)$, which is known via the \Am, we have been able to isolate the ``actual'' small parameter, i.e. $\boldsymbol{\lambda}(\bR,t)$. In all expressions, however, we find $\boldsymbol{\lambda}'(\bR,t)$, the matrix product of  $\dul{{\mathcal M}}^{-1}(\bR)$ and $\boldsymbol{\lambda}(\bR,t)$, which is a gauge-invariant quantity.

\section{Nuclear Hamiltonian}\label{app: app 4}
We show in this section the procedure leading to the appearance of the position-dependent mass $\dul{{\mathcal M}}(\bR)$ in the nuclear evolution equation~(1) of the exact factorization. The action of the kinetic energy operator $\hat{\tilde T}_n=\sum_\nu[-i\hbar\nabla_\nu+\bA_\nu]^2/(2M_\nu)$ on the nuclear wave function $\chi(\bR,t)$ can be written in matrix form as
\begin{align}\label{eqn: tilde Tn on chi 1}
\hat{\tilde T}_n\chi = \frac{1}{2}\big[-i\hbar\ul\nabla+\ul A\big]^T\dul{M}^{-1}\big[-i\hbar\ul\nabla+\ul A\big]\chi.
\end{align}
Using the expression~(\ref{eqn: tdvp}) of the TDVP, we identify the following terms
\begin{align}
\hat{\tilde T}_n\chi =\frac{1}{2}\bigg[&\left(-i\hbar\ul\nabla\right)^T\dul{M}^{-1}\left(\dul{\mathcal I}-\dul{\mathcal A}\,\dul{\mathcal M}^{-1}\right)\left(-i\hbar\ul\nabla\right)\nonumber\\
&-\left(\dul{\mathcal A}\,\dul{\mathcal M}^{-1}\ul\lambda\right)^T\dul M^{-1}\dul{\mathcal M}\left(\dul{\mathcal M}^{-1}\ul\lambda\right)\nonumber\\
&+\left(\dul{\mathcal A}\,\dul{\mathcal M}^{-1}\ul\lambda\right)^T\dul M^{-1}\left(\dul{\mathcal A}\,\dul{\mathcal M}^{-1}\ul\lambda\right)\bigg]\chi.\label{eqn: term 3}
\end{align}
In the second line we have used the definition of $\ul\lambda$ to write $-i\hbar\ul\nabla\chi=\ul\lambda\chi$ and we have inserted the definition of the identity matrix in the form $\dul{\mathcal I}=\dul{\mathcal M}^{-1}\dul{\mathcal M}$. We recall the expression of the position-dependent mass matrix, $\dul{\mathcal M}=\dul M+\dul{\mathcal A}$, leading to the kinetic energy operator in the nuclear Hamiltonian~(1),
\begin{align}\label{eqn: tilde Tn on chi 2}
\hat{\tilde T}_n\chi=&\frac{1}{2}\left(-i\hbar\ul\nabla\right)^T\dul{\mathcal M}^{-1} \left(-i\hbar\ul\nabla\right)\chi-\frac{1}{2}\left(\dul{\mathcal M}^{-1}\ul\lambda\right)^T\dul{\mathcal A}\left(\dul{\mathcal M}^{-1}\ul\lambda\right)\chi,
\end{align}
where only the position-dependent mass appears. In the second term on the right-hand-side, we have used the property of the \Am~of being symmetric, thus $\dul{\mathcal A}^T=\dul{\mathcal A}$. We can now show that this second term is exactly canceled out by a second order contribution in the potential energy of the nuclear Hamiltonian. In fact, in the kinetic energy, the product of two factors containing $\ul\lambda'=\dul{\mathcal M}^{-1}\dul\lambda$ is fundamentally a second order quantity. Therefore, we analyze the potential energy up to within second order terms in the perturbation.

The nuclear Hamiltonian from the exact factorization, in Eq.~(1), contains $\epsilon(\bR,t)$, the time-dependent potential energy surface. Therefore, we shall study its expression in order to identify a kinetic-like contribution to balance the second term in Eq.~(\ref{eqn: tilde Tn on chi 2}). We write the expression of $\langle\Phi_\bR(t)|\hat H_{BO}|\Phi_\bR(t)\rangle_\br$ up to within second order terms, when the electronic wave function is expanded as
\begin{align}
\Phi_\bR(\br,t)=\varphi_\bR^{(0)}(\br)+\lambda'(t)\varphi_{\bR}^{(1)}(\br)+\lambda'^2(t)\varphi_{\bR}^{(2)}(\br).
\end{align}
We use here a simplified notation, also using the property that the only time-dependence in the electronic wave function appears via $\boldsymbol\lambda_\nu'(\bR,t)$. Using this form of the electronic wave function, we write
\begin{align}
\left\langle\Phi_\bR(t)\right|\hat H_{BO}\left|\Phi_\bR(t)\right\rangle_\br =& \,\epsilon_{BO}^{(0)}(\bR)+\lambda'^2(t)\left\langle \varphi_{\bR}^{(1)}\right|\hat H_{BO}\left|\varphi_{\bR}^{(1)}\right\rangle_\br\nonumber \\
&+\lambda'^2(t)\epsilon_{BO}^{(0)}(\bR)\left[\left\langle\varphi_{\bR}^{(2)}\right|\left.\varphi_\bR^{(0)}\right\rangle_\br+\left\langle\varphi_\bR^{(0)}\right|\left.\varphi_{\bR}^{(2)}\right\rangle_\br\right]+\mathcal O(\lambda^3),\label{eqn: second order potential}
\end{align}
and, by using the partial normalization condition up to within second order,
\begin{align}
\left\langle\varphi_{\bR}^{(0)}\right|\left.\varphi_\bR^{(0)}\right\rangle_\br+\lambda'^2(t)&\left\langle\varphi_\bR^{(1)}\right|\left.\varphi_{\bR}^{(1)}\right\rangle_\br+\lambda'^2(t)\left\langle\varphi_{\bR}^{(2)}\right|\left.\varphi_\bR^{(0)}\right\rangle_\br+\lambda'^2(t)\left\langle\varphi_\bR^{(0)}\right|\left.\varphi_{\bR}^{(2)}\right\rangle_\br=1,
\end{align}
we find
\begin{align}
\left\langle\varphi_{\bR}^{(2)}\right|\left.\varphi_\bR^{(0)}\right\rangle_\br+\left\langle\varphi_\bR^{(0)}\right|\left.\varphi_{\bR}^{(2)}\right\rangle_\br =- \left\langle\varphi_\bR^{(1)}\right|\left.\varphi_{\bR}^{(1)}\right\rangle_\br,
\end{align}
since the normalization condition is already satisfied at zero-th order. We insert this result in Eq.~(\ref{eqn: second order potential}) to obtain
\begin{align}
\left\langle\Phi_\bR(t)\right|&\hat H_{BO}\left|\Phi_\bR(t)\right\rangle_\br = \epsilon_{BO}^{(0)}(\bR)+\lambda'^2(t)\left\langle \varphi_{\bR}^{(1)}\right|\hat H_{BO}-\epsilon_{BO}^{(0)}(\bR)\left|\varphi_{\bR}^{(1)}\right\rangle_\br+\mathcal O(\lambda'^3).
\end{align}
In the second term on the right-hand-side we identify the \Am~and we thus write
\begin{align}
\left\langle\Phi_\bR(t)\right|\hat H_{BO}\left|\Phi_\bR(t)\right\rangle_\br &=\epsilon_{BO}^{(0)}(\bR)+\sum_{\nu,\nu'}\sum_{i,j}\frac{1}{2}\lambda_{\nu i}'(\bR,t)\mathcal A^{ij}_{\nu\nu'}(\bR)\lambda_{\nu'j}'(\bR,t),\label{eqn: A L L 1}\\
&=\epsilon_{BO}^{(0)}(\bR)+
\frac{1}{2}{\ul{\lambda}'}^T(\bR,t)\dul{{\mathcal A}}(\bR)\ul{\lambda}'(\bR,t) \label{eqn: A L L 2}
\end{align}
where Eq.~(\ref{eqn: A L L 2}) is a rewriting of Eq.~(\ref{eqn: A L L 1}) in matrix form. Inserting the expression of $\ul{\lambda}'(\bR,t)$ in terms of $\ul{\lambda}(\bR,t)$ given in Eq.~(\ref{eqn: matrix relation between the parameters in app}), we can express the second term of Eq.~(\ref{eqn: A L L 2}) as
\begin{align}
{\ul{\lambda}'}^T(\bR,t)\dul{{\mathcal A}}(\bR)\ul{\lambda}'(\bR,t)=\left[\dul{{\mathcal M}}^{-1}(\bR)\ul{\lambda}(\bR,t)\right]^T\dul{{\mathcal A}}(\bR)\left[\dul{{\mathcal M}}^{-1}(\bR)\ul{\lambda}(\bR,t)\right],\nonumber
\end{align}
which exactly cancels the second term on the right-hand-side of Eq.~(\ref{eqn: tilde Tn on chi 2}). The nuclear Hamiltonian of Eq.~(13) is thus derived,
\begin{align}\label{eqn: Hn with M(R)}
\hat H_n=\frac{1}{2}\left(-i\hbar\ul\nabla\right)^T\dul{\mathcal M}^{-1}(\bR)\left(-i\hbar\ul\nabla\right)+ E(\bR).
\end{align}
The potential energy is time-independent and contains the BO energy, from the first term in Eq.~(\ref{eqn: A L L 2}), and an additional contribution, according to
\begin{align}
E(\bR)=\epsilon_{BO}^{(0)}(\bR)+\sum_{\nu=1}^{N_n}\frac{\hbar^2}{2M_\nu}\left\langle \nabla_\nu\varphi_\bR^{(0)}\right|\left.\nabla_\nu\varphi_\bR^{(0)}\right\rangle_\br.
\end{align}
It is worth noting that the first order contribution to the time-dependent potential $\epsilon(\bR,t)$ is zero, thus only $\epsilon^{(0)}(\bR)$, the zeroth order term, appears as potential energy in the nuclear Hamiltonian of Eq.~(13). This statement has been already proven in Ref.~\cite{Scherrer_JCP2015} using the definition in Eq.~(4) and the expression of the electronic wave function up to within first order terms in the perturbation. The second term on the right-hand-side is referred to as Born-Huang diagonal correction in the applications proposed in the paper. Among the second order contributions to the potential energy (it appears at the order $\mu^4$ in Eq.~(6)), only this term beyond $\epsilon_{BO}^{(0)}(\bR)$ will be included in the calculations, due to the fact that at this stage the theory does not allow us to efficiently compute higher order terms.

The correspondence principle of quantum mechanics enables us to determine the classical nuclear Hamiltonian as
\begin{align}\label{eqn: classical hamiltonian}
H_n = \frac{1}{2}\ul P^T\dul{\mathcal M}^{-1}(\bR) \ul P + E(\bR)
\end{align}
where $\ul P=\dul{\mathcal M}(\bR)\dot{\ul R}$ is the nuclear momentum.

\section{Electronic mass and the \texorpdfstring{$\boldsymbol{\mathcal A}$-}~matrix}\label{app: app 5}
The derivation of Eq.~(15) uses the property of the BO electronic wave function of being invariant under a translation of the coordinate reference system, namely $\varphi_{\bR'}^{(0)}(\br')=\varphi_{\bR}^{(0)}(\br)$ with $\bR'=\bR_1',\ldots,\bR_{N_n}'=\bR_1+\eta \boldsymbol\Delta,\ldots,\bR_{N_n}+\eta \boldsymbol\Delta$ and analogously for $\br'$. Notice that $\boldsymbol\Delta$ is a three-dimensional vector and that all positions, electronic and nuclear, are translated of the same amount $\eta \boldsymbol\Delta$. Translational invariance~\cite{Rauk_CP1993} means
\begin{align}
0 = \frac{\partial\varphi_{\bR'}^{(0)}(\br')}{\partial\eta} &=\sum_{i=x,y,z}\left[\sum_{\nu=1}^{N_n}\frac{\partial\varphi_{\bR'}^{(0)}(\br')}{\partial R_{\nu i}'}\frac{\partial R_{\nu i}'}{\partial\eta}+\sum_{k=1}^{N_{el}}\frac{\partial\varphi_{\bR'}^{(0)}(\br')}{\partial r_{k i}'}\frac{\partial r_{k i}'}{\partial\eta}\right]\nonumber\\
&=\sum_{i=x,y,z}\Delta_i\left[\sum_{\nu=1}^{N_n}\frac{\partial\varphi_{\bR'}^{(0)}(\br')}{\partial R_{\nu i}'}+\sum_{k=1}^{N_{el}}\frac{\partial\varphi_{\bR'}^{(0)}(\br')}{\partial r_{k i}'}\right] \nonumber\\
&=\boldsymbol\Delta\cdot\left[\sum_{\nu=1}^{N_n}\nabla_\nu\varphi_{\bR'}^{(0)}(\br')+\sum_{k=1}^{N_{el}}\nabla_k\varphi_{\bR'}^{(0)}(\br')\right],\label{eqn: translational invariance}
\end{align}
which is valid for all values of $\boldsymbol\Delta$. Identifying $\nabla_k$ as the position representation of the momentum operator $\hat{\mathbf p}_k$ corresponding to the $k$-th electron (divided by $-i\hbar$), which can be written also as
\begin{align}\label{eq:commutator}
 \hat{\mathbf p}_k = \frac{im}{\hbar} \left[\hat{H}, \hat{\mathbf r}_k\right] = \frac{im}{\hbar} \left[\hat{H}_{BO}, \hat{\mathbf r}_k\right],
\end{align}
and projecting the two terms in square brackets in Eq.~(\ref{eqn: translational invariance}) onto $\varphi_{\bR,\nu i}^{(1)}(\br)$, from Eq.~(\ref{eqn: nacvs in phi 1}),
\begin{align}
\sum_{\nu=1}^{N_n}\left\langle\varphi_{\bR,\nu' i}^{(1)}\right|\left.\hbar\nabla_\nu\varphi_{\bR}^{(0)}\right\rangle_\br=\frac{m}{\hbar}\sum_{k=1}^{N_{el}}\left\langle\varphi_{\bR,\nu' i}^{(1)}\right|\left[\hat{H}_\text{BO}, \hat{\mathbf r}_k\right]\left|\varphi_{\bR}^{(0)}\right\rangle,
\end{align}
we identify the \Am~on the left-hand-side and, for each Cartesian component $j$, we write
\begin{align}
\sum_{\nu=1}^{N_n} \mathcal A_{\nu'\nu}^{ij}(\bR) = -\frac{m}{\hbar}\sum_{k=1}^{N_{el}}\left\langle\varphi_{\bR,\nu' i}^{(1)}\right|  \left[\hat{H}_{BO}, \hat{r}_{k j}\right] \left|\varphi_{\bR}^{(0)}\right\rangle_\br.\label{eqn: APT}
\end{align}
From the term on the right-hand-side we derive the expression of the atomic polar tensor (APT). First of all we write explicitly the commutator and we use Eq.~(\ref{eqn: first order electronic eqn}) to obtain
\begin{align}
\left\langle\varphi_{\bR,\nu' i}^{(1)}\right|\left[\hat{H}_{BO}, \hat{r}_{k j}\right] \left|\varphi_{\bR}^{(0)}\right\rangle_\br&=\int d\br\,\varphi_{\bR,\nu' i}^{(1)}(\br)\left[\hat H_{BO}-\epsilon^{(0)}_{BO}(\bR)\right]r_{kj}\varphi_{\bR}^{(0)}(\br)\\
&=-\hbar\int d\br \left(\partial_{\nu'i}\,\varphi_{\bR}^{(0)}(\br)\right)r_{kj}\varphi_{\bR}^{(0)}(\br),
\end{align}
then we identify the expectation value of the electronic dipole moment operator over the BO wave function in the following expression
\begin{align}\label{eqn: APT 2}
\partial_{\nu'i}\sum_{k=1}^{N_{el}}\int d\br\,\varphi_{\bR}^{(0)}(\br) r_{kj}\varphi_{\bR}^{(0)}(\br) = \frac{1}{e}\partial_{\nu'i}\left\langle \hat \mu^{(el)}_{j}(\bR)\right\rangle_{BO}.
\end{align}
The derivative with-respect-to the $i$-th Cartesian component, relative to the $\nu'$-th nucleus, of the $j$-th Cartesian component of the electronic dipole moment is the definition of the electronic contribution to the APT~\cite{Person1974} $\mathcal{P}_{ij}^\nu(\bR)$. This leads to the relation~\cite{Stephens1990, Rauk_CP1993}
\begin{align}\label{eqn: sum rule app}
\sum_{\nu,\nu'=1}^{N_n} \mathcal A_{\nu'\nu}^{ij}(\bR) =\sum_{\nu=1}^{N_n}\frac{m}{e}\mathcal{P}_{ij}^\nu(\bR) = mN_{el}\delta_{ij},
\end{align}
when we further sum over the index $\nu$. This result states that when the \Am~is summed up over all nuclei it yields the total electronic mass of the complete system. In Eqs.~(13) and~(\ref{eqn: Hn with M(R)}) this means that the mass effect of the electrons is completely taken into account by the position-dependent mass corrections to the nuclear masses within the order of the perturbation considered here.

\section{Separation of the center of mass}\label{app: app 6}
We introduce the coordinate transformation
\begin{align}
\bR_1' &= \boldsymbol{\mathcal R}_{\textrm{CoM}}= \frac{1}{M_{tot}}\left[\sum_{\nu=1}^{N_n}M_\nu\bR_\nu + m\sum_{k=1}^{N_{el}}\left\langle\hat{\mathbf r}_k\right\rangle_{BO}\right] \label{eqn: CoM app}\\
\bR_\nu' &= \bR_\nu-\bR_1 \quad\mbox{with }\nu\geq2,
\end{align}
with the position of the center of mass (CoM) defined in Eq.~(\ref{eqn: CoM}) and $M_{tot}=\sum_{\nu}M_\nu+mN_{el}$ the total mass of the system. Such coordinate transformation is applied to the kinetic and potential energy terms in the nuclear Hamiltonian~(\ref{eqn: Hn with M(R)}). Since we have to evaluate the gradient of $\chi$, we have to compute the Jacobian matrix of the transformation from Cartesian to internal coordinates. The Jacobian is a ($3N_n\times3N_n$) matrix, whose elements are
\begin{align}\label{eq:jacobian-elements}
 J_{\nu\nu'}^{ij} = \frac{\partial R^\prime_{\nu i}}{\partial R_{\nu' j}} =  \left\{
  \begin{array}{l l}
   \frac{1}{M_{tot}} \left(M_{\nu'} \delta_{ij}  + \frac{m}{e}\mathcal{P}^{\nu'}_{ji} \right)  & \text{if } \nu=1 \\
    -\delta_{1\nu'}\delta_{ij} +\delta_{\nu\nu'}\delta_{ij} & \text{if } \nu\geq2
  \end{array}\right.
\end{align}
with $\mathcal{P}^{\nu'}_{ji}$ the electronic APT of Eq.~(\ref{eqn: APT 2}). It can be proved with some simple, but tedious, algebra that the determinant of the Jacobian is unity. In Eq.~(\ref{eqn: Hn with M(R)}) we replace $\ul{\nabla}$ with $\ul{\nabla}'$ according to
\begin{align}
\left(-i\hbar\ul{\nabla}\right)^T\dul{\mathcal M}^{-1}\left(-i\hbar\ul{\nabla}\right)=\left[\dul J^T\left(-i\hbar\ul{\nabla}'\right)\right]^T\dul{\mathcal M}^{-1}\left[\dul J^T\left(-i\hbar\ul{\nabla}'\right)\right]=\left(-i\hbar\ul{\nabla}'\right)^T\left(\dul J\,\dul{\mathcal M}^{-1}\dul J^T\right)\left(-i\hbar\ul{\nabla}'\right)
\end{align}
where the position-dependent mass in the last term on the right-hand-side depends on $\bR'$, namely
\begin{align}\label{eqn: mass in new coordinates}
\dul{{\mathcal M}}^{-1}(\bR')=\dul J\, \dul{{\mathcal M}}^{-1}(\bR)\dul J^T.
\end{align}
We rewrite the Jacobian matrix as the sum of two terms, $\dul J^{CoM}$ and $\dul J^{int.}$: the first three rows of $\dul J^{CoM}$ are the same as $\dul J$, thus given by Eq.~(\ref{eq:jacobian-elements}) for $\nu=1$, i.e. $\left(J^{CoM}\right)_{\nu\nu'}^{ij}=\delta_{\nu1}J_{\nu\nu'}^{ij}$, with each row composed by $3N_n$ entries, all other elements of $\dul J^{CoM}$ are zeros; the first three rows of $\dul J^{int.}$ are zero and the remaining $3(N_n-1)$ rows are the same as $\dul J$, thus given by the second expression in Eq.~(\ref{eq:jacobian-elements}). We now introduce the operator $\dul{\mathcal{T}}$, defined as $\mathcal{T}_{\nu\nu'}^{ij}= \delta_{ij}\delta_{\nu'1}$, and we notice that the product of the position-dependent mass matrix and $\dul{\mathcal{T}}$ yields 
\begin{align}\label{eqn: MT}
\dul{{\mathcal M}}(\bR)\, \dul{\mathcal{T}} &= M_{tot}\left[\dul J^{CoM}\right]^T,
\end{align}
as we will now prove. First of all, we recall the expression of the position-dependent mass matrix,
\begin{align}
 \mathcal M_{\nu\nu'}^{ij}(\bR) = M_\nu\delta_{\nu\nu'}\delta_{ij} + \mathcal A_{\nu\nu'}^{ij}(\bR),
\end{align}
then we write the matrix product with $\dul{\mathcal{T}}$ as the sum of their components, namely
\begin{align}
 \sum_{j=x,y,z}\sum_{\nu'=1}^{N_n}\mathcal M_{\nu\nu'}^{ij}(\bR) \mathcal{T}_{\nu'\nu''}^{jk} = \left(M_\nu \delta_{ik} + \frac{m}{e}\mathcal{P}^{\nu}_{ik}(\bR)\right) \delta_{\nu''1} = M_{tot}\left[\delta_{\nu''1}J_{\nu\nu''}^{ik}\right]^T,
\end{align}
where we used the sum rule of Eq.~(\ref{eqn: sum rule}) in the first equality and Eq.~(\ref{eq:jacobian-elements}) in the second. We identify the term in square brackets in the last equality as $\dul J^{CoM}$. Further relations that will be used below are
\begin{align} \label{eq:jacobian_com_proj}
 \dul{J}^{CoM}\,\dul{\mathcal{T}} &= \left(\begin{array}{cc}
 \dul{I}^{(3)} & \dul 0\\
 \dul 0 & \dul 0\\
\end{array}\right) \\
 \dul{J}^{int.}\,\dul{\mathcal{T}} &= \dul 0. \label{eq:jacobian_rest_proj}
\end{align}
Eq.~(\ref{eqn: mass in new coordinates}) is written by introducing the two components, $CoM$ and $int.$, of the Jacobian as
\begin{align}\label{eqn: M(R')}
\dul{{\mathcal M}}^{-1}(\bR')= &\,
 \dul{J}^{CoM}\,\dul{{\mathcal M}}^{-1}(\bR)\left[\dul{J}^{CoM}\right]^T+\dul{J}^{int.}\dul{{\mathcal M}}^{-1}(\bR)\left[\dul{J}^{int.}\right]^T\nonumber\\
&+\dul{J}^{int.}\dul{{\mathcal M}}^{-1}(\bR) \left[\dul{J}^{CoM}\right]^T+\dul{J}^{CoM}\dul{{\mathcal M}}^{-1}(\bR) \left[\dul{J}^{int.}\right]^T.
\end{align}
Using Eq.~(\ref{eqn: MT}), the first term on the right-hand-side can be rewritten as
\begin{align}
\dul{J}^{CoM} \dul{{\mathcal M}}^{-1}(\bR) \left[\dul{J}^{CoM}\right]^T=\frac{1}{M_{tot}}\dul{J}^{CoM}\dul{{\mathcal M}}^{-1}(\bR)\dul{{\mathcal M}}(\bR)\dul{\mathcal{T}}=\frac{1}{M_{tot}}\dul{J}^{CoM}\dul{\mathcal{T}},
\end{align}
and from Eq.~(\ref{eq:jacobian_com_proj}) we obtain
\begin{align}
 \frac{1}{2}\left(-i\hbar\ul{\nabla}'\right)^{T}\left[\dul{J}^{CoM}\dul{{\mathcal M}}^{-1}(\bR) \left[\dul{J}^{CoM}\right]^T \right]\left(-i\hbar\ul{\nabla}'\right)= \frac{\hat P_{\mathrm{CoM}}^2}{2M_{tot}}.
\end{align}
A similar procedure, which uses Eq.~(\ref{eq:jacobian_rest_proj}), is employed to show that the cross terms (second and third terms on the right-hand-side) in Eq.~(\ref{eqn: M(R')}) do not contribute to the kinetic energy. Therefore, the final result reads
\begin{align} \label{eq:separated-hamiltonian}
\hat H_n = \frac{\hat P_{\mathrm{CoM}}^2}{2M_{tot}} +  \frac{1}{2}\left(-i\hbar\ul{\nabla}'\right)^{T}\dul{{\mathcal M}}(\bR')\left(-i\hbar\ul{\nabla}'\right) +E(\bR').
\end{align}

\section{Numerical details of the O-H-O model}\label{app: app 7}
A model of a proton involved in a one-dimensional hydrogen bond like $\textrm O-\textrm H-\textrm O$ is considered~\cite{Borgis_CPL2006}, with potential
\begin{align}\label{eqn: test potential}
V(r,R) = &\,D\left[e^{-2a\left(\frac{R}{2}+r-d\right)}-2e^{-a\left(\frac{R}{2}+r-d\right)}+1\right]\nonumber\\
&+Dc^2\left[e^{-\frac{2a}{c}\left(\frac{R}{2}-r-d\right)}-2e^{-\frac{a}{c}\left(\frac{R}{2}-r-d\right)}\right]+Ae^{-BR}-\frac{C}{R^6}.
\end{align}
Here $r$ indicates the position of the proton measured from the center of the $\textrm O-\textrm O$ bond and $R$ stands for the $\textrm O-\textrm O$ distance. The chosen parameters of the Morse potential are $D=60$~kcal/mol, $d=0.95$~\AA, $a=2.52$~\AA$^{-1}$; $c=0.707$ makes the potential for the proton asymmetric, mimicking a strong $\textrm O-\textrm H-\textrm O$ bond. The other parameters are $A=2.32\times10^{5}$~kcal/mol, $B=3.15$~\AA$^{-1}$ and $C=2.31\times 10^{4}$~kcal/mol/\AA$^{6}$. The full Hamiltonian of the system involves $V(r,R)$ and the kinetic energies of the oxygen atoms and of the proton, namely
\begin{align}
\hat H(r,R_{\mathrm O^-},R_{\mathrm O^+})&=\sum_{\nu=+,-}\frac{-\hbar^2\nabla^2_{{\mathrm O^\nu}}}{2M_{\mathrm O^\nu}} + \frac{-\hbar^2\nabla^2_r}{2M_{\mathrm H}}+\hat V\left(r,R_{\mathrm O^-},R_{\mathrm O^+}\right)\label{eqn: full hamiltonian}\\
&=\sum_{\nu=+,-}\frac{-\hbar^2\nabla^2_{{\mathrm O^\nu}}}{2M_{\mathrm O}} + \hat H_{BO}\left(r,R_{\mathrm O^-},R_{\mathrm O^+}\right),\label{eqn: BO hamiltonian}
\end{align}
where $\hat V$, according to Eq.~(\ref{eqn: test potential}), depends only on the distance between the oxygen atoms, $R=|R_{\mathrm O^-}-R_{\mathrm O^+}|$.

In the static calculations, the adiabatic states have been computed by diagonalizing the BO Hamiltonian in Eq.~(\ref{eqn: BO hamiltonian}) on a spatial grid $400\times400$. The eigenvalues of the full Hamiltonian in Eq.~(\ref{eqn: full hamiltonian}) are determined using a Gaussian quadrature method with 20 points for $R$, the distance between the two heavy atoms, and 34 for $r$, the displacement of the proton from the CoM of the heavy atoms. When the Hamiltonian with position-dependent dressed masses is used for computing the eigenvalues, $R$ is again the distance between the twxo heavy atoms. In this case, as described in the text, the BO approximation has been introduced before separating the CoM motion and the eigenvalues of the Hamiltonian in internal coordinates (indicated by the prime symbols in Eq.~(18)) have been computed.

In the dynamics we use the three coordinates, i.e. $R_{\mathrm{O}^+}$, $R_{\mathrm{O}^-}$ and $r=r_{\mathrm{H}}$, in order to test the conservation of the position of the CoM. The results in the paper are shown for the same number of periods in all cases, using: the velocity-Verlet algorithm to integrate the classical equations, with a time-step 1~fs; the Crank-Nicolson~\cite{Nicolson_PCPS1947} algorithm for the proton (quantum) equation in Ehrenfest, with a time-step $10^{-4}$~fs; the Euler algorithm if the force depends on the velocity (see Eq.~(\ref{eqn: force with velocity})) with time-step $0.0625$~fs, where the stability of the integration has been tested based on the energy conservation. The position of the proton is estimated as the expectation value of the position operator on the proton wave function at the instantaneous $\textrm O-\textrm O$ geometry.

\subsection{Calculation of the \texorpdfstring{$\boldsymbol{\mathcal A}$-}~matrix}\label{app: calc a-matrix}
We have computed the \Am~using density functional perturbation theory~\cite{Scherrer_JCTC2013, Baroni_PRL1987,Baroni-2001,Gonze-1995b, Putrino-2000, Watermann-2014} as described in Ref.~\cite{Scherrer_JCP2015} and checked that the sum rule of Eq.~(\ref{eqn: sum rule}) is satisfied. The numerical scheme has been implemented in the electronic structure package CPMD~\cite{CPMD}. Calculations have been performed using Troullier-Martins~\cite{Troullier-1991} pseudo-potentials in the Becke-Lee-Yang-Parr~\cite{Becke1988, Lee1988} (BLYP) approximation of the exchange-correlation kernel. The molecular geometry is the equilibrium geometry at the BLYP level, employing the aug-cc-pVTZ basis set~\cite{Kendall1992} in the Gaussian electronic structure program~\cite{g09}.

Table~\ref{tab:a mat h2} shows the results for the H$_2$ molecule. Remember that the \Am~is a ($3N_n\times 3N_n$) matrix, with blocks
\begin{align}
\left(
\begin{array}{@{\vrule height 10.5pt depth7pt  width0pt}c|c}
\left( \ul{\mathcal A}_{\mathrm H_1\mathrm H_1}\right)^{ij} & \left( \ul{\mathcal A}_{\mathrm H_1\mathrm H_2}\right)^{ij} \\
\hline
\left( \ul{\mathcal A}_{\mathrm H_2\mathrm H_1}\right)^{ij} & \left( \ul{\mathcal A}_{\mathrm H_2\mathrm H_2}\right)^{ij}
\end{array}
\right)
\end{align}
and indices $i,j$ running over the Cartesian components $x,y,z$, so each block is a ($3\times3$) matrix. 
\begin{table}[ht!]
 \begin{center}
 \bgroup
 \def\arraystretch{0.75}
 \begin{tabular}{@{\vrule height 10.5pt depth4pt  width0pt}l|rrr|rrr}
   \hline
  \hline
  & \multicolumn{3}{c}{Hydrogen 1} & \multicolumn{3}{c}{Hydrogen 2} \\
  \hline
  \hline
  \multirow{3}{*}{Hydrogen 1}
  &   0.553 &          &          &    0.446 &          &           \\
  &         &    0.553 &          &          &    0.446 &           \\
  &         &          &    0.868 &          &          &    0.131  \\
  \hline
  \multirow{3}{*}{Hydrogen 2}
  &         &          &          &    0.553 &          &           \\
  &         &          &          &          &    0.553 &           \\
  &         &          &          &          &          &    0.868  \\
   \hline
   \hline  
 \end{tabular}
 \caption{Diagonal elements of the \Am~in the case of the H$_2$ molecule (oriented along z-axis).}
 \label{tab:a mat h2}
 \egroup
 \end{center}
\end{table}
The sum rule in Eq.~(\ref{eqn: sum rule}) reads, in this case,
\begin{align}
\sum_{\nu,\nu'=1}^{N_n} &\mathcal A_{\nu'\nu}^{xx}(\bR) = \Big[\left(\mathcal A_{\mathrm H_1\mathrm H_1}\right)^{xx}+\left(\mathcal A_{\mathrm H_1\mathrm H_2}\right)^{xx}+\left(\mathcal A_{\mathrm H_2\mathrm H_1}\right)^{xx}+\left(\mathcal A_{\mathrm H_2\mathrm H_2}\right)^{xx}\Big]=1.998\simeq2
\end{align}
and similarly for the other Cartesian components. This result is obtained by summing the entries of the matrix in Table~\ref{tab:a mat h2}, and we find indeed the total electronic mass ($m=1, N_{el}=2$) of the system as  expected from Eq.~(\ref{eqn: sum rule}).

In the case of the H$_2$O molecules the use of non-local pseudo-potentials poses additional technical complications that we discuss here. The BO Hamiltonian in Eq.~(\ref{eq:commutator}) contains a potential energy term corresponding to the pseudo-potential, namely
\begin{align}
 \hat{H}_{BO} = \hat{T}_e + \hat{V}_\text{loc} + \hat{V}_\text{nl}.
\end{align}
$\hat{V}_\text{nl}$, the non-local part of the pseudo-potential, does not commute with the position operator~\cite{Pickard-2003} thus we have to take into account such correction when deriving the sum rules of Eq.~(\ref{eqn: sum rule}).

\begin{table*}[hb!]
 \begin{center}
 \bgroup
 \def\arraystretch{0.75}
 \begin{tabular}{@{\vrule height 10.5pt depth4pt  width0pt}l|rrr|rrr|rrr}
  \hline
  \hline
  \multicolumn{1}{c}{\phantom{\Big[}$\tilde{\mathcal A}$}& \multicolumn{3}{c}{Oxygen} & \multicolumn{3}{c}{Hydrogen 1} & \multicolumn{3}{c}{Hydrogen 2} \\
  \hline
  \hline
  \multirow{3}{*}{Oxygen}
   & 6.500 &          &          &    0.190 &          &   -0.091 &    0.190 &          &    0.091  \\
   &       &    6.323 &          &          &    0.415 &          &          &    0.415 &           \\
   &       &          &    5.989 &   -0.172 &          &    0.382 &    0.172 &          &    0.382  \\
  \hline
  \multirow{3}{*}{Hydrogen 1}
   &       &          &          &    0.658 &          &    0.287 &    0.020 &          &    0.082  \\
   &       &          &          &          &    0.314 &          &          &    0.037 &           \\
   &       &          &          &          &          &    0.527 &   -0.082 &          &    0.052  \\
  \hline
  \multirow{3}{*}{Hydrogen 2}
   &       &          &          &          &          &          &    0.658 &          &   -0.287  \\
   &       &          &          &          &          &          &          &    0.314 &           \\
   &       &          &          &          &          &          &          &          &    0.527  \\
  \hline
  \hline
 \end{tabular}
 \caption{Local part of the \Am, i.e. $\tilde{\mathcal A}$, in the case of an isolated H$_2$O molecule (in xz-plane oriented along z-axis).}
  \label{tab:a_matrix_correction_h20 1}
 \egroup
 \end{center}
\end{table*}
The evaluation of Eq.~(\ref{eqn: APT}) using only local pseudo potentials in the commutator in Eq.~(\ref{eq:commutator}) gives rise to the \Am~contribution due to the \textit{local} pseudo-potentials, in the following termed \textit{local part} of the \Am. The local part of the \Am~is indeed symmetric and has positive diagonal elements. However, it does not satisfy the sum rule of Eq.~(\ref{eqn: sum rule}), as shown in Table~\ref{tab:a_matrix_correction_h20 1}. In order to correct for this error, we can calculate the correction due to the full commutator where also the effect of the non-local pseudo-potential is included. For all $3N_n$ nuclear coordinates, labeled by the indices $i,\nu$, we obtain a commutator for each Cartesian component $j$. The correction hence gives rise to a $(3N_n\times3)-$dimensional matrix $\Delta^{ij}_\nu$, i.e. the non-local contribution to the electronic APT. However, the appropriate dimension of the matrix to be used to correct the \Am~should be ($3N_n\times3N_n$), as the \Am~itself. Unfortunately, there is no protocol that allows us to match the dimensions of the two matrices, i.e. the \Am~and the correction matrix, based on some physical properties. Therefore, we develop such protocol according to the following prescription. The correction matrix is denoted $\Delta \mathcal A^{ij}_{\nu\nu}$ and is shown in Table~\ref{tab:a_matrix_correction_h20 2}. 
\begin{table*}[h!]
 \begin{center}
 \caption{Symmetrized correction $\Delta \mathcal A$ in the case of an isolated H$_2$O molecule (in xz-plane oriented along z-axis).}
  \label{tab:a_matrix_correction_h20 2}
 \bgroup
 \def\arraystretch{0.75}
 \begin{tabular}{@{\vrule height 10.5pt depth4pt  width0pt}l|rrr|rrr|rrr}
  \hline
  \hline
  \multicolumn{1}{c}{\phantom{\Big[}$\Delta \mathcal A$}& \multicolumn{3}{c}{Oxygen} & \multicolumn{3}{c}{Hydrogen 1} & \multicolumn{3}{c}{Hydrogen 2} \\
  \hline
  \hline
  \multirow{3}{*}{Oxygen}
  &   -0.395 &          &          &          &          &          &          &          &           \\
  &          &   -0.493 &          &          &          &          &          &          &           \\
  &          &          &   -0.415 &          &          &          &          &          &           \\
  \hline
  \multirow{3}{*}{Hydrogen 1}
  &          &          &          &   -0.112 &          &   -0.081 &          &          &           \\
  &          &          &          &          &   -0.093 &          &          &          &           \\
  &          &          &          &          &          &   -0.133 &          &          &           \\
  \hline
  \multirow{3}{*}{Hydrogen 2}
  &          &          &          &          &          &          &   -0.112 &          &    0.081  \\
  &          &          &          &          &          &          &          &   -0.093 &           \\
  &          &          &          &          &          &          &          &          &   -0.133  \\
   \hline
   \hline
 \end{tabular}
 \egroup
 \end{center}
\end{table*}
We add the symmetric part of the correction $\Delta^{ij}_\nu$, in Table~\ref{tab:a_matrix_correction_h20 3}, corresponding to each nucleus to the diagonal parts of the blocks of the \Am, i.e.~for each nucleus $\nu$:
\begin{align}
 \Delta \mathcal A^{ij}_{\nu\nu} = \frac{1}{2}(\Delta_\nu^{ij}+\Delta_{\nu}^{ji}) \quad \forall \nu,i,j.
\end{align}
\begin{table*}[h!]
 \begin{center}
 \bgroup
 \def\arraystretch{0.75}
 \begin{tabular}{@{\vrule height 10.5pt depth4pt  width0pt}l|rrr|rrr|rrr}
  \hline
  \hline
  \multicolumn{1}{c}{\phantom{\Big[}$\Delta$}& \multicolumn{3}{c}{Oxygen} & \multicolumn{3}{c}{Hydrogen 1} & \multicolumn{3}{c}{Hydrogen 2} \\
  \hline
  \hline
  \multirow{3}{*}{Correction}
   &  -0.395 &          &          &   -0.112 &          &   -0.081 &   -0.112 &          &    0.081  \\
   &         &   -0.493 &          &          &   -0.093 &          &          &   -0.093 &           \\
   &         &          &   -0.415 &   -0.081 &          &   -0.133 &    0.081 &          &   -0.133  \\
 \hline
 \hline
 \end{tabular}
 \caption{Non-local pseudo-potential correction $\Delta$ in the case of an isolated H$_2$O molecule (in xz-plane oriented along z-axis).}
  \label{tab:a_matrix_correction_h20 3}
 \egroup
 \end{center}
\end{table*}
This correction leads to a correct sum rule for the \Am~whereas preserving all the known symmetry properties. We show the final result of this operation in Table~\ref{tab:a_matrix_correction_h20 4}. In the case of the water molecule we compute
\begin{align}
\sum_{\nu,\nu'=1}^{N_n} &\mathcal A_{\nu'\nu}^{xx}(\bR) =7.997\simeq8.
\end{align}
The sum rule~(\ref{eqn: sum rule}) yields a total mass of 8 ($m=1, N_{el}=8$), which is the number of electrons that are considered explicitly. The two $1s$ electrons of the oxygen atom are treated in the frozen core approximation.
\begin{table*}[ht!]
 \begin{center}
 \bgroup
 \def\arraystretch{0.75}
 \begin{tabular}{@{\vrule height 10.5pt depth4pt  width0pt}l|rrr|rrr|rrr}
  \hline
  \hline
  \multicolumn{1}{c}{\phantom{\Big[}$\mathcal A$}& \multicolumn{3}{c}{Oxygen} & \multicolumn{3}{c}{Hydrogen 1} & \multicolumn{3}{c}{Hydrogen 2} \\
  \hline
  \hline
  \multirow{3}{*}{Oxygen}
   &    6.105 &          &          &    0.190 &          &   -0.091 &    0.190 &          &    0.091  \\
   &          &    5.830 &          &          &    0.415 &          &          &    0.415 &           \\
   &          &          &    5.574 &   -0.172 &          &    0.382 &    0.172 &          &    0.382  \\
  \hline
  \multirow{3}{*}{Hydrogen 1}
   &          &          &          &    0.546 &          &    0.206 &    0.020 &          &    0.082  \\
   &          &          &          &          &    0.222 &          &          &    0.037 &           \\
   &          &          &          &          &          &    0.394 &   -0.082 &          &    0.052  \\
  \hline
  \multirow{3}{*}{Hydrogen 2}
   &          &          &          &          &          &          &    0.546 &          &   -0.206  \\
   &          &          &          &          &          &          &          &    0.222 &           \\
   &          &          &          &          &          &          &          &          &    0.394  \\
 \hline
 \hline
 \end{tabular}
 \caption{Corrected \Am~in the case of an isolated H$_2$O molecule (in xz-plane oriented along z-axis). The sum rule in Eq.~(\ref{eqn: sum rule}) yields, for the three Cartesian components $xx$, $yy$ and $zz$, $mN_{el}=7.998, 8.006, 7.993$, respectively.}
  \label{tab:a_matrix_correction_h20 4}
 \egroup
 \end{center}
\end{table*}

\subsection{Normal mode analysis}
It is easy to prove that given a Lagrangian of the form 
\begin{align}
\mathcal{L}(\dot{\bR},\bR) = \frac{1}{2}{\dot{\ul{R}}}^T \dul{{\mathcal M}}(\bR) \dot{\ul{R}} - E(\bR),
\end{align}
the classical Hamiltonian of Eq.~(\ref{eqn: classical hamiltonian}) can be derived as its Legendre-transform. Therefore, nuclear motion is classically governed by the Euler-Lagrange equation
\begin{align}\label{eqn: force with velocity}
\dul{{\mathcal M}}(\bR) \ddot{\ul{R}} = -\ul\nabla E(\bR)-\frac{1}{2}\dot{\ul{R}}^{T} [\ul\nabla\,\dul{{\mathcal M}}(\bR)] \dot{\ul{R}}.
\end{align}
This classical equation of motion is integrated using the Euler algorithm as described in Appendix~\ref{app: app 7}. If (i) we use internal coordinates, since the free motion of the CoM can be separated as in Eq.~(\ref{eq:separated-hamiltonian}), (ii) we introduce the harmonic approximation of $E(\bR)$ and (iii)  we neglect the velocity-dependent term, we obtain
 \begin{align}
 \ddot{\ul{R}} = -[\dul{{\mathcal M}}^{-1}(\bR_0) \dul{K}(\bR_0)]\ul{R},
 \end{align}
 with $\dul{K}$ the Hessian matrix computed from the ground state electronic potential. The term in square brackets is evaluated at the equilibrium geometry $\bR_0$. The diagonalization of the matrix in square brackets yields corrected $\nu+\Delta \nu$ frequencies, as $\Delta \nu$ includes the effect of electrons that follow the motion of the nuclei non-adiabatically, namely not instantaneously.

\end{document}